%% file: main.tex
\documentclass[sigconf,9pt]{acmart}

\copyrightyear{2023}
\acmYear{2023}
\setcopyright{rightsretained}
\acmConference[SPAA '23]{Proceedings of the 35th ACM Symposium on Parallelism in Algorithms and Architectures}{June 17--19, 2023}{Orlando, FL, USA}
\acmBooktitle{Proceedings of the 35th ACM Symposium on Parallelism in Algorithms and Architectures (SPAA '23), June 17--19, 2023, Orlando, FL, USA}
\acmDOI{10.1145/3558481.3591071}
\acmISBN{978-1-4503-9545-8/23/06}

\usepackage{booktabs}   
\usepackage{subcaption} 
\usepackage{caption}
\usepackage{colortbl}
\usepackage{style}
\usepackage{bigstrut}
\usepackage{microtype}
\usepackage{tabularx}
\usepackage{multirow}
\usepackage{multicol}
\usepackage{rotating}
\usepackage{lipsum}
\usepackage{balance}
\usepackage{mfirstuc}
\usepackage{titlecaps}
\usepackage{listings}
\usepackage{float}
\usepackage[rightcaption]{sidecap}
\usepackage[compact]{titlesec}

\titleformat{\section}{\Large\bfseries}{\thesection}{1em}{}
\titleformat{\subsection}{\large\bfseries}{\thesubsection}{1em}{}
\titleformat{\subsubsection}[runin]{\bfseries}{\thesubsubsection}{1em}{}

\usepackage{xcolor}
\usepackage[font=small,labelfont=bf,skip=2pt]{caption}

\input{macro.tex}
\renewcommand\footnotetextcopyrightpermission[1]{} 

\begin{document}

\fancyhead{}
\title{High-Performance and Flexible Parallel Algorithms for \\Semisort and Related Problems}


\author{Xiaojun Dong}
\affiliation{%
  \institution{University of California, Riverside}
  \country{}
}
\email{xdong038@ucr.edu}

\author{Yunshu Wu}
\affiliation{%
  \institution{University of California, Riverside}
  \country{}
}
\email{ywu380@ucr.edu}

\author{Zhongqi Wang}
\affiliation{%
  \institution{University of Maryland, College Park}
  \country{}
}
\email{zqwang@umd.edu}

\author{Laxman Dhulipala}
\affiliation{%
  \institution{University of Maryland, College Park}
  \country{}
}
\email{laxman@umd.edu}

\author{Yan Gu}
\affiliation{%
  \institution{University of California, Riverside}
  \country{}
}
\email{ygu@cs.ucr.edu}

\author{Yihan Sun}
\affiliation{%
  \institution{University of California, Riverside}
  \country{}
}
\email{yihans@cs.ucr.edu}


\input{abstract}




\maketitle

\input{intro}
\input{prelim}
\input{algo-new}



\input{exp}

\input{conclusion}

\clearpage
\bibliographystyle{abbrv}

\input{semisort.bbl}
\clearpage
\appendix
\clearpage
\input{appendix.tex}

\end{document}
\endinput

%% file: macro.tex
  \newcommand{\fname}[1]{\textsc{#1}}
  \newcommand{\vname}[1]{\mathit{#1}}

  \newcommand{\variablename}[1]{{\texttt{#1}}}

\definecolor{forestgreen}{rgb}{0.13, 0.55, 0.13}

  \newcommand{\semisort}{{semisort}\xspace}
  \newcommand{\Semisort}{{Semisort}\xspace}
  \newcommand{\SMS}{SMS}
  
  \newcommand{\semisortequal}{{semisort$_=$}\xspace}
  \newcommand{\semisortless}{{semisort$_<$}\xspace}
  \newcommand{\semisortequali}{semisort-i$_=$\xspace{}}
  \newcommand{\semisortlessi}{semisort-i$_<$\xspace{}}
  \newcommand{\histogram}{{histogram}\xspace}
  \newcommand{\collectreduce}{{collect-reduce}\xspace}
  \newcommand{\stepname}[1]{{\textit{#1}}}
  \newcommand{\sampling}{{\stepname{Sampling and Bucketing}}\xspace}
  \newcommand{\counting}{{\stepname{Blocked Distributing}}\xspace}
  
  \newcommand{\placing}{\stepname{Placing}\xspace}
  \newcommand{\refining}{\stepname{Local Refining}\xspace}
  
  \newcommand{\heavyhash}{H}
  \newcommand{\heavytable}{heavy table}
  \newcommand{\B}{l}
  
  \newcommand{\countm}{C}
  \newcommand{\presum}{X}
  \newcommand{\countmatrix}{counting matrix\xspace}
  \newcommand{\userhash}{{user hash function}\xspace}
  
  \newcommand{\basecase}{\alpha}

  \newcommand{\record}{record}
  \newcommand{\key}{\mathit{key}}
  \newcommand{\BGSS}{GSSB\xspace}
  \newcommand{\gssb}{GSSB\xspace}
  \newcommand{\block}{subarray\xspace}
  \newcommand{\bucket}{bucket\xspace}
  \newcommand{\nheavy}{n_H} 
  \newcommand{\nlight}{n_L} 
  \newcommand{\hashrange}{\kappa} 

  \newcommand{\semieq}{semisort\_equal}
  \newcommand{\semile}{semisort\_less}
  \newcommand{\counts}{\vname{count}}
  \newcommand{\offsets}{\vname{offsets}}
  \newcommand{\comp}{\fname{comp}}

\newcommand{\modelop}[1]{\texttt{#1}}
\newcommand{\forkins}{\modelop{fork}}

\newcommand{\thread}{thread}

\newcommand{\cas}{{\texttt{compare\_and\_swap}}}

\newcommand{\matrixtrans}{\intercal}

\newcommand{\bool}{\emph{Boolean}}

\newcommand{\ifconference}{{{\ifx\fullversion\undefined}}}

\makeatletter
\def\dfnt@space@setup{%
\dfnt@preskip=\parskip
  \dfnt@postskip=0pt}
\makeatother

\newtheoremstyle{exampstyle}
{.05in} 
{.05in} 
{} 
{.5em} 
{\sc \bfseries} 
{.} 
{.5em} 
{} 
\theoremstyle{exampstyle} 
\theoremstyle{exampstyle} 

\makeatletter
\renewenvironment{proof}[1][\proofname]{\par
\vspace{-\topsep}
\pushQED{\qed}%
\normalfont
\topsep0pt \partopsep0pt 
\trivlist
\item[\hskip\labelsep
      \itshape
  #1\@addpunct{.}]\ignorespaces
}{%
\popQED\endtrivlist\@endpefalse
\addvspace{3pt plus 3pt} 
}

 \crefname{section}{Sec.}{Sec.}
 \crefname{theorem}{Thm.}{Thm.}
 \crefname{lemma}{Lem.}{Lem.}
 \crefname{corollary}{Col.}{Col.}
 \crefname{table}{Tab.}{Tabs.}
 \crefname{algorithm}{Alg.}{Algs.}
 \Crefname{table}{Tab.}{Tabs.}

%% file: abstract.tex
\begin{abstract}
Semisort is a fundamental algorithmic primitive widely used in the design
and analysis of efficient parallel algorithms.
It takes input as an array of records and a function extracting a
\emph{key} per record, and reorders them so that records with equal keys are
contiguous.
Since many applications only require collecting equal values, but not fully
sorting the input, semisort is broadly applicable, e.g., in string algorithms,
graph analytics, and geometry processing, among many other domains.
However, despite dozens of recent papers that use semisort in their theoretical
analysis and the existence of an asymptotically optimal parallel semisort
algorithm, most implementations of these parallel algorithms choose to
implement semisort by using comparison or integer sorting in practice, due to potential performance
issues in existing semisort implementations.

In this paper, we revisit the semisort problem, with the goal of achieving
a high-performance parallel semisort implementation with a flexible interface.
Our approach can easily extend to two related problems, \emph{histogram} and \emph{collect-reduce}.
%
%
Our algorithms achieve strong speedups in practice, and importantly, outperform
state-of-the-art parallel sorting and semisorting methods
for almost all settings we tested, with varying input sizes, distribution, and key types. 
%
%
We also test two important applications with real-world data, and show that our
algorithms improve the performance over existing approaches.
We believe that many other parallel algorithm implementations can be
accelerated using our results.
\end{abstract}

%% file: intro.tex
\section{Introduction}
\label{sec:intro}
The \defn{\semisort{}} problem takes as input an array of \record{s} with
associated keys, and returns a reordered array such that all \record{s} with
identical keys are contiguous.
Importantly, the problem does not require all keys to appear in sorted order in
the output, nor all \record{s} with distinct keys to be sorted.
Several other important and widely-applicable problems are closely related to
semisort, such as the \defn{\histogram{}} problem that counts the number of
occurrences of each key, and the more general \defn{\collectreduce{}} problem
that computes the aggregate ``sum'' for each key, based on all the \record{s}.
The ``sum'' function can be defined based on any
associative (sometimes also commutative) combine function (e.g., addition or maximum).
Semisort, \histogram{}, and \collectreduce{} are all widely used in different
areas, but are often referred to using different names, e.g.,
\variablename{groupBy}/\variablename{aggregation} in databases~\cite{muller2015cache,do2023efficient}, \variablename{frequency} in data analytics
applications, \variablename{reduceByKey}/\variablename{groupByKey} in RDD in
Spark~\cite{zaharia}, the \variablename{shuffle} step in the MapReduce
paradigm~\cite{dean08}, and others~\cite{henriksen2020compiling}.
%
As an example of the applicability of these problems, consider a database of sales
receipts keeping the information of each sold lineitem. Useful operations
to analyze trends in this data include quickly gathering lineitems from the
same branch together (\semisort{}), counting the number of sold items in each
month (\histogram{}), and obtaining the total sale of lineitems of each brand
(\collectreduce{}).

Semisorting was first studied as a theoretical problem by Valiant to
efficiently simulate parallel machine models (e.g., the PRAM) with other
models~\cite{Valiant91}.
Sequentially, it is easy to semisort in $O(n)$ time using a hash table,
and theoretically-efficient parallel algorithms are also known~\cite{gu2015top}.
Today, in contrast to its initial development as a theoretical tool for machine
simulations, semisort is widely used in the design and analysis of efficient
and practical parallel algorithms, for example for graph
analytics~\cite{dhulipala2017,dhulipala2019low,dhulipala2020semi,dhulipala2020connectit,gbbs2021,blelloch2016parallelism,blelloch2021read,dong2023provably,dhulipalathesis,quanquanthesis,
blelloch2016parallel,acar2019parallel,dong2021efficient,liu2021parallel,qiu2021lightne,tseng2019batch,shun2020practical,shi2021parallel,acar2020parallel,anderson2020work,shi2020parallel},
geometry
problems~\cite{blelloch2018geometry,wang2021fast,wang2020theoretically,GuThesis,shen2022many},
sequence algorithms and many
others~\cite{blelloch2020optimal,blelloch2020parlaylib,tangwongsan2019parallel,kang2021processing,sunthesis,kaler2021parad,gu2022parallel,yang2022optimal,ellert2020lcp}.
However, there is a \emph{disconnect between theory and practice} in these parallel applications.
In all of the above-mentioned papers, semisort is only used in theoretical analysis to
obtain better bounds by the theoretically-efficient parallel semisort algorithm~\cite{gu2015top},
but is not used in practical
implementations of these algorithms.
In particular, for the papers that implement and evaluate their parallel
algorithms, a comparison sort (usually the samplesort
in~\cite{blelloch2020parlaylib,blelloch2010low,axtmann2017place}) is used.
Although semisorting is asymptotically simpler than sorting, semisorting
is avoided in practice in favor of sorting the data.


The only known parallel semisort algorithm and implementation is by Gu et
al.~\cite{gu2015top} in 2015 (the \gssb{} algorithm), with $O(n)$ expected work
(number of operations) and space, and $O(\log n)$ span (longest dependencies) \whp{}~\cite{blelloch2020optimal}.
Despite the asymptotic guarantees, the algorithm has not been
widely used in practice for a few reasons.
First, the algorithm uses many random accesses and is I/O-unfriendly since it
heavily uses hash tables.
%
%
Second, the algorithm \emph{interface} also incurs performance overhead.
Specifically, the algorithm assumes the records are associated with
\emph{hashed keys} with no duplicates rather than arbitrary keys (more details in \cref{sec:gssb}).
This assumption requires additional steps to hash the original keys and resolve
collisions subsequently, which may take time comparable to \semisort{} itself.
Although none of these issues increase the asymptotic bounds, they both
contribute to performance slowdowns that are hard to avoid in a faithful
implementation.
Hence, the semisort implementation
in~\cite{gu2015top} is not faster than many recent sorting
algorithms~\cite{blelloch2020parlaylib,axtmann2022engineering,obeya2019theoretically} in
practice.
Meanwhile, the \gssb{} algorithm is not stable or (internally-)deterministic
(i.e., the result may depend on runtime scheduling)
due to the use of parallel hash tables.

\defn{In this paper, we revisit the \semisort{} problem, with the goal of achieving
a high-performance parallel semisort implementation with a flexible interface.
}
%
%
We propose new parallel \semisort{} algorithms that are efficient regarding work, I/O and space usage.
Our flexible interface for \semisort{} can also be extended to support
efficient and parallel \histogram{} and \collectreduce{}.
Our algorithm takes any key type $K$, and a user-defined hash
function $h:K\mapsto [1,\dots,n^\hashrange]$ to map keys to integers.
In principle, the only extra information we need is an \emph{equality test}
$=_K:K\times K\mapsto \text{\bool}$.
We observe that in most use cases, the key type also supports a \emph{less-than}
test $<_K:K\times K\mapsto \text{\bool}$ to determine a total ordering, which
can be used to improve the performance. We refer to the general
\semisort{} algorithm (only $=_K$ supported) and the version with $<_K$ as
\semisortequal{} and \semisortless{}, respectively.

Our algorithm builds on the strengths of \gssb{}~\cite{gu2015top},
but substantially redesigns several components to overcome the
existing performance issues of the \gssb{} algorithm.
\gssb{} works in three steps (we review more details in \cref{sec:gssb}).
It first uses samples to determine the heavy (frequent) and light (infrequent) keys,
and constructs \bucket{s} for them based on estimated sizes from the samples.
Each heavy key will be in a separate bucket,
while multiple light keys can be grouped into the same bucket.
The algorithm then scatters all records to their buckets by placing each record to a random slot in their bucket (or linear probe when the slot is occupied).
Lastly, the algorithm refines each light bucket by radix sorting (the hashed keys) in each light bucket.
\hide{We then split the input into \block{s}, and deal with the \block{s} in parallel.
In each \block{}, we sequentially count the number of \record{s} falling into each \bucket{}, and store the results in a matrix $C$.
This matrix yields the (exact) needed size of each \bucket{}.
Finally, we place the \records{} in each \block{} to the output array sequentially, but process all \block{s} in parallel.
This is again given by the matrix $C$, which provides the offset of each \record{} in the corresponding \bucket{}.}
The main issue in \gssb is that the scatter phase is implemented using
a parallel hash table, which causes excessive random memory access,
some space overhead, instability, and non-determinism.

To avoid the use of a parallel hash table, we propose an idea inspired by the I/O-efficient parallel samplesort~\cite{blelloch2010low}: when constructing
buckets and scattering records, we split the input into consecutive \block{s}, use auxiliary arrays to count the appearance of each bucket in each \block, and distribute the records in each \block based on the counts.
This approach enables a cache-friendly access pattern to the input, allows us to obtain the exact size of each bucket, and is stable and race-free (and thus deterministic).
%
%
However, since samplesort and semisort are quite different,
using the idea in~\cite{blelloch2010low} does not directly enable high-performance.
The challenge lies in choosing the best parameters for the number of heavy and light buckets.
At a high-level, the samplesort in ~\cite{blelloch2010low} creates a bucket for every sampled key.
However, using too many buckets increases the size of the auxiliary counting array,
which can greatly increase memory accesses.
On the other hand, having more buckets is useful to improve parallelism, since each bucket can be processed independently in parallel.
Specifically for semisort, we also wish to create more heavy buckets because heavy keys do not need to be refined and are easier to process.

To create the heavy and light buckets in the best way,
we propose novel algorithmic ideas for semisort.
First, we control the parameters to keep the number of buckets small, so that the auxiliary arrays fit in cache.
This avoids excessive memory access to the auxiliary arrays (see more details about the auxiliary arrays in \cref{sec:counting,fig:semisort}).
Second, we deal with each light bucket \emph{recursively} in parallel.
To enable efficient recursive calls, we carefully design optimizations to avoid extra space in recursive calls.
Our new approach saves the main memory accesses for the auxiliary arrays,
and more interestingly, identifies more heavy keys than \gssb{} with different degrees of ``heaviness'' using recursions.
The ``relatively heavy'' keys in each light bucket can be identified and handled more efficiently and improve the overall performance.

\hide{
Our idea is to control the parameters to fit the auxiliary arrays in cache, to avoid the excessive memory accesses to the auxiliary arrays.
However, note that fitting the arrays in cache means to reduce the number of heavy and light buckets and thus significantly increase the bucket sizes (see more details about the auxiliary arrays in \cref{sec:counting,fig:semisort}) and sacrifices parallelism.
To tackle this, we then recursively \semisort{} the light buckets.
Our new approach saves the main memory accesses for the auxiliary arrays, and more interestingly, identifies \emph{more} heavy keys than the GSSB algorithm with different degrees of ``heaviness''. The ``relatively heavy'' keys in each light bucket can be identified and handled more efficiently.
We note that although the high-level idea sounds simple, it involves some details, such as avoiding extra space needed in recursive calls. 
%
By combining all approaches, with more details given in \cref{sec:alg}, our new algorithms achieve much better performance, and is also stable and race-free.
}

\input{figures/heatmap.tex}

In addition to algorithmic improvement for performance, we also redesigned the algorithm interface.
Our algorithm directly takes the input with any key type, a user-defined hash function $h$, and an equality test (or less-than for \semisortless{}).
This avoids the additional pre- and post-processing in \gssb{}.
Due to the more flexible interface and algorithm design, our \semisort{} algorithm can be easily extended to \histogram{} and \collectreduce{} (see \cref{sec:histogram}).

We tested our algorithms on a variety of benchmarks, with different core counts, input sizes, key lengths, and distribution patterns (uniform, exponential, and Zipfian).
We summarize our results as a heatmap in \cref{fig:heatmap}.
We also test two applications: graph transposing (reordering graph edge lists), and $k$-gram on English texts.
Both our \semisortequal{} and \semisortless{} algorithms achieve high performance on almost all tests.
For example, on $10^9$ input data with 64-bit keys and 64-bit values over 15 distributions,
our algorithm is 3.4$\times$ faster than the \gssb{} algorithm, and at least 1.35$\times$ faster than the best of the previous algorithms, both on average (geometric mean).
Our algorithms also consistently perform well on the two applications with real-world data.
In all but one application-input combination we tested in this paper, our algorithm is the fastest.
Our code is publicly available~\cite{semisortcode}, and we plan to integrate it into ParlayLib~\cite{blelloch2020parlaylib}.


\hide{
Surprisingly, although semisort and its related family of problems are important
and widely-used in algorithm design and analysis, there have been very few
\laxman{empirical?} studies on this problem. One possible reason for the lack of
study of this problem is that the problem itself is trivial in the sequential
setting. Specifically, a hash table can be used to solve any of the three
problems in linear time and space.  However, in parallel, finding an efficient
solution is more challenging.
Consider \semisort{ing} as an example.  One could consider a solution using
a parallel hash table with chaining using linked lists, but observe that this
sequentializes all operations when all \record{s} have the same key, which
completely destroys parallelism. Another simple solution is to create a sufficiently
large array for each cell in the hash table for chaining, and when putting a
\record{} into a hash table cell, scatter it to a random position of the
associated array.  However, this requires to know the number of occurrences for
each key to allocate appropriate size of memory. It is almost equally hard as
the \semisort{ing} \xiaojun{semisort or semisorting?} \laxman{imo `semisort'}
(this is exactly the \histogram{} problem). Another effective sequential
solution for \semisort{} is to hash all keys into range $[1,\dots,n^k]$ ($n$ is
the input size), and use an integer sort, but parallel integer sort itself is
a notoriously challenging open problem~\cite{}.

Theoretically this is not a problem since hashing is asymptotically bounded by
\semisort{}. However, in practice, this can also incur significant work and
space, even comparable to the \semisort{} algorithm itself.
As a result, most existing parallel software~\cite{} still choose to use an
$O(n\log n)$-work sorting algorithm when a \semisort{} primitive is needed, in
most cases the \emph{sample sort}, as sample sort is well-accepted as the
fastest parallel sorting algorithm because of I/O-efficiency.

probability \laxman{Isn't the span $O(\log n)$ in the BF model?}. The main idea
of this algorithm is to use sampling to  \emph{heavy keys}, which are keys that
appear more often, and \emph{light keys} otherwise, and to deal with them with
separate mechanisms.
%
%
\laxman{This would be a good time to state the wide use of this result in
parallel algorithms papers.}
Despite the asymptotic theoretical guarantees, the algorithm has not been
widely adopted in practice for a few reasons.  Firstly, the algorithm
relies  heavily on hash tables and random scattering \laxman{is random
scattering clear?}, which incur large amounts of random accesses, making the
algorithm less cache-friendly.
For the same reason, as these techniques usually requires $(1+\epsilon)\times$
more space than needed (for some constant $\epsilon>0$), the algorithm can
consume significant extra space.
In fact, their experiments use at least $5.5\times n$ space, where $n$ is the
input size\footnote{Their paper did not report memory usage, but it is not hard
to compute from the parameters provided in the paper.}. This overhead in space
increases memory footprint and can significantly harm performance.
Lastly, the algorithm assumes the input to be \emph{hashed keys}, which are
always integers. They excluded the time for hashing and resolving collisions.
Theoretically this is not a problem since hashing is asymptotically bounded by
\semisort{}. However, in practice, this can also incur significant work and
space, even comparable to the \semisort{} algorithm itself.
As a result, most existing parallel software~\cite{} still choose to use an
$O(n\log n)$-work sorting algorithm when a \semisort{} primitive is needed, in
most cases the \emph{sample sort}, as sample sort is well-accepted as the
fastest parallel sorting algorithm because of I/O-efficiency.
}

%% file: figures/heatmap.tex
\begin{figure}
  \centering
  \vspace{-.5em}
  \includegraphics[width=.95\columnwidth]{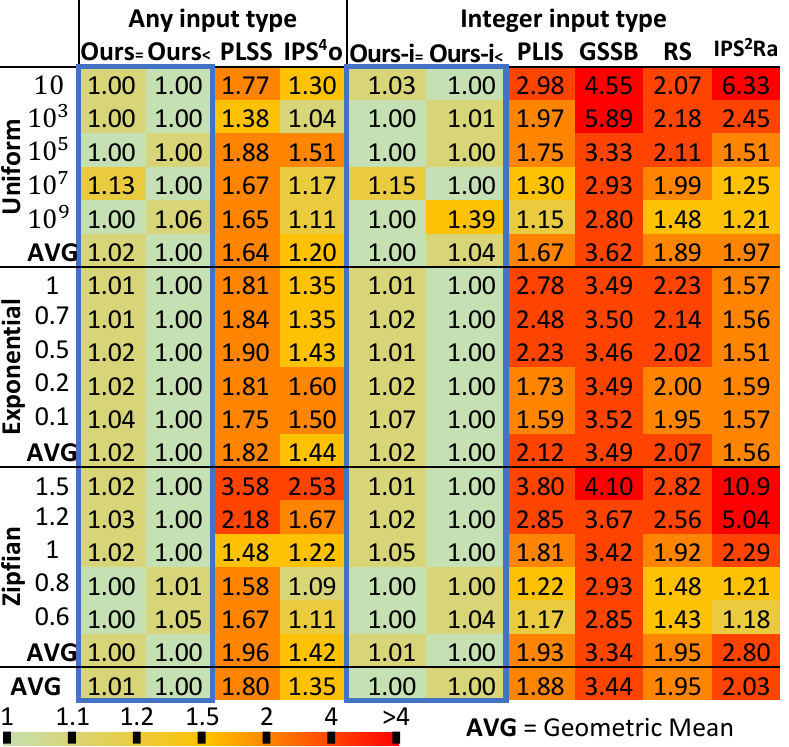}
  \caption{\small\textbf{Heatmap of the relative performance of implementations normalized to the fastest in each test (each row).}
  $n=10^9$. 64-bit keys and 64-bit values.
  The parameters in exponential distributions are multiplied by $10^{4}$.
  The algorithm names and details are introduced in \cref{tab:baseline}.
  }\label{fig:heatmap}
  \vspace{-1.5em}
\end{figure} 

%% file: prelim.tex
\section{Preliminaries}\label{sec:prelim}

\subsection{Problem Definitions}
Given a sequence of \defn{\record{s}} from a universe $U$, we define a key function
$\key:U\mapsto K$ to define the \defn{key} for each record, where $K$ is the
key type.
%
We define $=_K$
as the \defn{equality test} on $K$. When applicable, we use $<_K$ as the
\defn{less-than test} on $K$.  Given a sequence of \record{s} $A$, its key
function $\key_A$, and the
equality test $=_K$ on $K$, the \defn{\semisort{}} problem is to reorder the
\record{s} in $A$ to $A'$ such that all \record{s} with the same key are
contiguous in $A'$.  We also require the user to provide a family of hash
functions $h:K \mapsto [1, \dots , n^{\hashrange}]$, for some constant
$\hashrange\ge 1$. We call $h(\cdot)$ the \emph{\userhash}.

Given $A$, $\key_A$, $=_K$, and $h_A$, the \defn{\histogram{}} problem
is to emit an array of key-value pairs $G$ consisting of the unique keys of
$A$, with the value for each key equal to the number of times it appears in
$A$.
The \defn{\collectreduce{}} function takes the same arguments as \semisort{}
and two additional functions: a \emph{map} function $M : U \rightarrow E$,
and a \emph{reduce} monoid $(\oplus_E,I_E)$.
The map function maps a record to a value of some type $E$.
The reduce operation $\oplus_E : E \times E \rightarrow E$ combines values of type $E$ with identity $I_E$.
The collect-reduce function returns the array of key-value pairs $R\in K\times E$ consisting of the unique keys
of $A$, with the value associated with each key $k$ equal to $\oplus_{r \in S_k} M(r)$,
where $S_k=\{r\in A~|~ \key_A(r) =_K k\}$.
Note that \histogram{} can be expressed as \collectreduce{}
where the map function is the constant function $1$, and the monoid is $(+, 0)$.
With clear context, we drop the subscripts for these operations and functions.


\subsection{Computational Models and Other Notations}
We use the work-span (or work-depth) model for fork-join parallelism with
binary forking to analyze parallel algorithms~\cite{CLRS,blelloch2020optimal},
which is recently used in many papers on parallel
algorithms~\cite{agrawal2014batching,blelloch2010low,blelloch1999pipelining,BlellochFiGi11,dinh2016extending,xu2022efficient,xu2020parallel,blelloch2018geometry,dhulipala2020semi,BBFGGMS18,blelloch2020randomized,gu2021parallel,ahmad2021low,goodrich2021atomic,dhulipala2022hierarchical}.
%
We assume a set of \thread{}s that share a common memory.
A process can \forkins{} two child software \thread{s} to work in parallel.
When both children complete, the parent process continues.
The \defn{work} of an algorithm is the total number of instructions and
the \defn{span} (depth) is the length of the longest sequence of dependent instructions in the computation.
We can execute the computation using a randomized work-stealing
scheduler~\cite{BL98,ABP01} in practice.

To measure the memory access cost in an algorithm, we use the classic I/O model~\cite{AggarwalV88,Frigo99}.
We assume a two-level memory hierarchy.
The processor is connected to the cache of size $M$, and the cache is connected to an infinite-size main memory.
Both cache and main memory are divided into blocks (cachelines) of size $B$, so there are $M/B$ cachelines in the cache.
The CPU can only access the memory on blocks resident in the cache and it is free of charge.
We assume an optimal offline cache replacement policy to transfer the data between the cache and the main memory, and a unit cost for each cacheline load and evict.
The \defn{I/O cost} of an algorithm is the total cost to execute this algorithm on this model.

We say that a sorting/semisorting algorithm is \defn{stable} if the output preserves the relative order among equal keys from the input order, and otherwise we say that the algorithm is unstable.

We say an algorithm is \defn{race-free} when no two concurrent operations in the algorithm can access the same memory access and at least one of them is a write~\cite{CLRS}.
A race-free algorithm is (internally) deterministic~\cite{blelloch2012internally},
and has many advantages including ease of reasoning about the code, verifying correctness, debugging, and analyzing the performance.
In our algorithms, all operations in the algorithm are deterministic once we fix the random seed.

We use $O(f(n))$ \emph{with high probability (\whp{})} in $n$ to mean $O(cf(n))$ with probability at least $1-n^{-c}$ for $c \geq 1$.

\subsection{The \BGSS{} \Semisort{} Algorithm}\label{sec:gssb}
We first review the existing \gssb{} \semisort{} algorithm~\cite{gu2015top}.
As mentioned, the practical performance of \BGSS{}
is limited due to its 
excessive random
memory accesses and restrictive interface. Our algorithm builds on the
strengths of the \gssb{}, while overcoming the aforementioned limitations.
The \BGSS{} algorithm assumes the input as a sequence of \emph{hashed keys} in
range $[0, \dots, n^{\hashrange}]$ for some constant $\hashrange\ge 1$, and
\semisort{s} the hashed keys.

\myparagraph{Sampling and Bucketing}.
This is a key technique in \BGSS to handle heavily duplicate keys.
\BGSS{} first selects a sequence $S$ of samples from the input
sequence $A$ with sample rate $p=O(1/\log n)$.
The samples will be used to give an initial partition of the \record{s} into buckets,
such that the same key always goes to the same bucket.
Based on the samples, the keys are divided into \defn{heavy keys} and \defn{light keys} otherwise.
We call the \record{s} with heavy (light) keys the \defn{heavy (light) \record{s}}.
The theory behind this idea is that if sufficient ($\Omega(\log n)$) samples for a key $k$ can be obtained,
one can estimate the frequency of $k$ (relatively) accurately.
We call them the \defn{heavy keys} or \defn{heavy records}.
Let $\nheavy$ be the number of heavy keys identified by the samples.
The algorithm will construct $\nheavy$ \defn{heavy buckets}, each for an
individual heavy key.
Meanwhile, a key $k$ with few ($o(\log n)$) samples are unlikely to appear many
times in the input, and we call them \defn{light keys} or \defn{light records}.
The light \record{s} are grouped into $\nlight=\Theta(n/\log^2
n)$ \defn{light \bucket{s}} by using the hashed value to randomly map
to one of the $\nlight$ buckets.
Our new algorithm will also use a similar technique to detect heavy (duplicate) keys, but with different parameters for better performance.

\input{tables/notations.tex}

\myparagraph{Size Estimation and Scattering}.
For a bucket with $s$ samples, \BGSS uses a \defn{size estimation} function $f(s)$  to
upper bound bucket size \whp{}.
The algorithm will allocate an array of size $(1+\epsilon) f(s)$ for this bucket for some constant $\epsilon>0$.
Then each \record{} is scattered to a random position in the corresponding bucket.
This is performed by using \cas{}, which atomically puts the \record{} into the position,
and re-picking another position using linear probing upon collisions or conflicts.
Our new algorithms do not use this approach.

\myparagraph{Local Sort and Pack.} After scattering, all the heavy keys are
collected in individual heavy \bucket{s}. Each light \bucket{} can contain more
than one light key type.
The records in a \bucket{} may not be contiguous due to the random scattering.
\gssb{} then uses a radix sort (on the
hashed keys) to refine light buckets (comparison sort is used in practice) and make them contiguous.
A \defn{packing
step} is needed for heavy \bucket{s} to put records in contiguous slots.
Our new algorithm also uses different approaches in this step.

\medskip
The main performance issue in \gssb{}
is the random access in the scatter phase---each record is assigned to
a random location, and has to retry if necessary. \BGSS hence
needs $O(n)$ random writes for the scattering phase, which is
I/O-inefficient. 
This also requires slightly more space (and thus memory footprint) since we need to ensure a load factor $c<1$.
We will show how to overcome this issue, as well as to make our new algorithms stable and race-free in \cref{sec:alg}.

Another major issue of GSSB is its interface.
GSSB assumes a collision-free hash function $h:K \mapsto [1, \dots , n^{\hashrange}]$ that maps arbitrary key types to random integers (hashed keys), and the algorithm (and implementation) directly semisort the hashed keys, which are random integers.
When using more realistic and practical hash functions with possible collisions, one has to perform preprocessing and postprocessing to deal with collisions.
While such pre/postprocessing do not asymptotically increase the cost of the algorithm in theory,
they can in practice incur significant time overheads
comparable to semisort itself ($O(n)$), and therefore make using semisort in applications prohibitively costly, relative to sorting.

%% file: tables/notations.tex
\begin{table}
  \centering \small
  \begin{tabular}{cl}
    \hline
    \multicolumn{2}{@{}l}{\textbf{Input:}}\\
    $A[1..n]$ & input array of \record{s} in universe $U$ \\
    $K$ & key type of \record{s}\\
    $\key(\cdot)$ & $\key:U\mapsto K$ extracts the key of a \record{} \\
    $=_K$ & (or $=$) equality test on keys \\
    $<_K$ & (or $<$) less-than test on keys \\
    $h(\cdot)$ & \userhash{}; $h:K\mapsto [0, n^{\hashrange}]$\\
    \multicolumn{2}{@{}l}{\textbf{Tunable Parameters:}}\\
    $\B$ & \block{} size\\
    $\basecase{}$ & base case threshold\\
    $\nlight=2^b$& number of light \bucket{s}\\
    \multicolumn{2}{@{}l}{\textbf{Other notations used in the algorithm and description:}}\\
    $n'$ & problem size of the current recursion\\
    $S$ & the set of samples. $|S|=\nlight \log n$\\
    $\nheavy$& number of heavy \bucket{s}, $\nheavy=O(\nlight)$\\
    $\heavyhash$ & \heavytable{}; Maps heavy keys to \bucket{} ids\\
    $\countm$ & \countmatrix{}\\
    $\presum$ & (column-major) prefix sum of $\countm$ \\
    \hline
  \end{tabular}
  \vspace{-.2in}
  \caption{\small \textbf{Notations and parameters used in our algorithms.
  }
  }\label{tab:notation}
\end{table}

%% file: algo-new.tex
\input{algo-semisort-new.tex}

\section{Our New Algorithms}\label{sec:alg}

In this section, we present our algorithms for \semisort{} and related problems.
We present the useful notations in \cref{tab:notation}.
For simplicity, we first focus on \semisort{} and then explain how to modify it to adapt to
\histogram{} and \collectreduce{} in \cref{sec:histogram}.

Our \semisort{} algorithm follows the same framework as \gssb{}, but employs
novel techniques to improve the performance for \emph{all the steps}.
Our new algorithm is \textbf{I/O-friendly, stable}, and \textbf{race-free}.
%
In contrast to \gssb{}, we do not require pre-hashing the keys.
Our algorithm directly handles input records of any
type, and extracts the hashed keys by applying the \userhash{} in the algorithm when
needed.
This generality in the interface also improves efficiency both in time and space---it avoids the pre- and post-processing, as well as the hash table to pre-hash keys and resolve collisions,
which can incur another $O(n)$ random reads and $O(n)$ extra space.
%
Our algorithm is \emph{stable}---all records with the same key will be kept in the same order in the output.
This feature is useful for \collectreduce{} and \histogram{} and increases
their generality, as discussed in \cref{sec:histogram}.
Our algorithm is also \emph{race-free}, which means no concurrent
writes are needed to any shared memory position.
This also makes our algorithms simple, practical,
and \emph{internally-deterministic} (i.e., the output does not depend on runtime scheduler). 

\input{figures/fig-semisort.tex}

We start by overviewing the high-level idea, and then present more
details in \cref{sec:sample,sec:counting,sec:refine}. 
We discuss how to support \histogram{} and \collectreduce{} in \cref{sec:histogram}.
In \cref{sec:para}, we present the theoretical analysis of our algorithm, and discuss the choices of parameters in theory and in practice.
Finally, we discuss the improvements of our algorithms over \gssb{} in \cref{sec:discussion-gssb}, and the comparison to existing sorting algorithms in \cref{sec:discussion-sort}.
\begin{enumerate}[leftmargin=*]
  \item \textbf{\sampling{}.} First, the algorithm performs sampling to find the heavy keys.
  Similar to \gssb{}, each heavy key uses an individual bucket, and multiple light keys share a \bucket{}.
  However, we pick a smaller number of buckets for a better overall memory-access pattern (see discussions in \cref{sec:discussion-gssb}). 

  \item \textbf{\counting{}.} Next, it counts the exact number of records in each bucket.
  Given the bucket sizes, the algorithm distributes input records to their associated buckets in an \emph{I/O-friendly} manner. 
      By performing \emph{exact} counting, the temporary
      arrays used are only of size $n$, and no parallel hash tables
      are necessary (as in \gssb{}).
      This distribution step makes the algorithm stable and race-free.

  \item \textbf{\refining{}.} After Step 2, the heavy keys are at
  their final positions in the heavy buckets.
  For light buckets, unlike \gssb{}, our algorithm
  \emph{recursively semisorts} them until the recursive input size is small enough (i.e., fitting in cache),
  at which point the keys are semisorted sequentially.
  This approach allows the algorithm to detect ``medium-level'' heavy keys in subsequent recursive rounds and also reduce the total number of I/Os.
\end{enumerate}

The pseudocode of our algorithm is given in \cref{alg:semisort}, and a running example is given in \cref{fig:semisort}.
Next, we introduce the details of each step and explain why our decisions improve the performance of our algorithm.
Since our algorithm uses recursive calls, we use $n$ as the input size of the original (top-level) problem, and use $n'$ as
the current subproblem size in the recursive call.

\subsection{Step 1: {\sampling{}}}\label{sec:sample}
The goal of the sampling and bucketing step is similar to \BGSS{}---we want to identify heavy and light keys, which decides the bucket id for each record.
Instead of having $O(n/\log^2 n)$ light buckets as in \gssb{},
we use the number of heavy and light buckets ($\nheavy$ and $\nlight$) as parameters.
We use $\nlight$ as a tunable parameter, and set the upper bound of $\nheavy$ accordingly as $O(\nlight)$.
We will later discuss in \cref{sec:para} about how to pick $\nlight$ to achieve the best practical performance.

To determine heavy keys, we take a sequence of samples $S$ of size $\Theta(\nlight\log n)$ by selecting each \record{} uniformly at random (Lines~\ref{line:sample}--\ref{line:step1:end} in \cref{alg:semisort}).
The light keys are keys appearing fewer than $\log n$ times, which indicates that their actual number of occurrences is small.
We group multiple light keys into one bucket based on the hash keys.
We create $\nlight$ light \bucket{s} by evenly partitioning the range of hashed keys (given by the \userhash{}) into $\nlight$ buckets.
For simplicity, we use $n_L=2^b$ as a power of 2, and the light bucket id of a key~$k$ is obtained by taking the last (least significant) $b$ bits in the hash value of $k$, i.e., the bucket id is $(h(k)\modulo n_L)$. 

The heavy keys are those appearing at least $\log n$ times in the samples; as in the analysis of \gssb{}, their actual number of occurrences is large \whp{}.
Given the sample size $|S|=\Theta(\nlight\log n)$, the number of heavy keys $\nheavy=O(\nlight)$.
We will create $\nheavy$ buckets with ids in $[\nlight,\nlight+\nheavy)$ (the first $\nlight$ buckets are for the light keys).
We use a sequential hash table $\heavyhash{}$ to store all heavy keys associated
with their \bucket{} ids, referred to as the \emph{\heavytable}, so that the
later steps can look up whether a key is heavy in constant work.

\hide{
The \emph{heavy keys} are the keys appearing at least $\log n$ times; as in the analysis of \gssb{}, we expect their actual number of occurrences to be large with high probability.
Using standard concentration inequalities, a heavy key must appear $\Omega(\log n)$ times in $S$ to enable an accurate estimation,
and the expected number of heavy keys $\nheavy=\Theta(\nlight)$.
We will create $\nheavy$ buckets with ids in $[\nlight,\nlight+\nheavy)$ (the first $\nlight$ buckets are for the light keys, see discussions below).
We use a hash table to store all heavy keys $\heavyhash{}$ associated
with their \bucket{} ids, referred to as the \emph{\heavytable}, so that the
subsequent steps can look up whether a key is heavy in constant work.

We now discuss the part for light buckets, which is relatively easy.
Recall that each heavy key is assigned an individual bucket, and multiple light keys share a \bucket{}.
We create $\nlight$ light \bucket{s} by evenly partitioning the range of hashed keys (given by the \userhash{}) into $\nlight$ buckets.
For simplicity, we assume $n_L=2^b$ as a power of 2, and the light bucket id of a key~$k$ is obtained by taking the last (least significant) $b$ bits in the hash value of $k$, i.e., the bucket id is $(h(k)\modulo n_L)$. 
}

\hide{
\myparagraph{Count Sort.}
We can assign an id to each light key by using the $b$ lowest significant bits of its hash value.
Assume we identify $n_H$ heavy keys and use $n_L=2^b$ ids to distinguish the light keys, then all keys can be partition into $n_H+n_L$ buckets (\cref{line:bucketid}).
In this step, we first (virtually) divide the input sequence into $\sqrt{n}$ blocks, and allocate a $\counts[\cdot]$ array to indicate the number of keys in each bucket in each block.
We can count the keys of different blocks in parallel.
Within each block, the $\counts[\cdot]$ array is computed in sequential.
Then the $\offsets[\cdot]$ of each block can be computed by using a scan function,
where $\offsets[\cdot]$ denotes the starting position of each bucket of each block when the keys are partitioned.
Lastly, we partition the keys and return the $\offsets[\cdot]$ of each light bucket.
}

\subsection{Step 2: \counting{}}\label{sec:counting}

Unlike \gssb{}, which uses a \emph{scattering} step to place records to random positions in the buckets, our algorithm
uses a more I/O-friendly and space-efficient approach.
The goal of this step is to count the \emph{exact} number of \record{s} in each \bucket{}, and distribute all records to the associated buckets into contiguous slots.
Since we have the exact count, we only need an array $T$ of size $n$ for all the buckets, making our algorithm space-efficient.
This distribution step makes our algorithm stable and race-free.

Our idea is inspired by recent sorting algorithms~\cite{axtmann2022engineering,blelloch2010low,obeya2019theoretically,merrill2017cub}.
We first partition the sequence evenly into $n'/\B$ \block{s}, each with $\B$ records (recall that $n'$
is the current subproblem size). 
We then process all the \block{s} in parallel (\cref{line:compute-c}), but sequentially within each
individual \block{} (\cref{line:seq-loop-1}).
We count the number of \record{s} in each bucket in a $(n'/l)\times (\nlight+\nheavy)$
matrix $\countm$, which is referred to as the \emph{\countmatrix}.  In particular,
$\countm_{ij}$ is the number of \record{s} in \block{} $i$ falling into
\bucket{} $j$.
To do this, within each \block{} $i$, our algorithm determines which bucket each key $k$
belongs to using the \textsc{GetBucketId} function (\cref{line:bucketid}).
This function first looks up the \heavytable{} $\heavyhash$ to check if $k$ is
a heavy key, and if so, it obtains the bucket id $j$ from $\heavyhash$.
Otherwise, the bucket id of a light key $k$ is simply given by
$j=h(k)\modulo n_L$.
We then increment the corresponding cell in $\countm_{ij}$ by one (\cref{line:cplus}).

We then distribute all records in the input to their corresponding buckets,
using the information in $\countm$.
To do so, we compute the offset per subarray per bucket as a \emph{prefix array} $\presum$ that has the same size as $C$.
Array $\presum$ can be computed using the prefix sum of $C$, but in the column-major order (\cref{line:offset}, see an illustration in \cref{fig:semisort}).
After the prefix array $\presum$ is computed, we once again process each subarray and move each record to its corresponding bucket (\cref{line:distribute-start}--\ref{line:distribute-end}) in the temporary array $T$.
This step takes $O(1)$ work per \record{}---we use $O(1)$ work to decide which bucket a record is in, and after that, we move the record and increment the offset counter in~$\presum$.

We noticed that when picking the appropriate parameters, our approach is much faster than the corresponding step in \gssb{} in practice, mainly due to smaller memory footprint and fewer memory accesses.
We will later show the analysis in \cref{sec:para}.


\subsection{Step 3: \refining}\label{sec:refine}

After the previous step, we have all heavy keys stored contiguously in their corresponding heavy buckets, which are also their final positions in $T$.
Light keys are still unsorted.
We work on each light \bucket{} in parallel by recursively semisorting each of them.
We stop recursing and switch to the base case when the bucket size is small enough and fits in cache, which is decided by the parameter~$\alpha$ (\cref{line:basecase}).
For our experiments with input sizes ranging from $10^8$--$10^9$, we typically need one more level of recursion before reaching the base case, if most of the keys are light keys.
Since the base-case size fits in cache, the time to semisort the base cases is small.
We provide two solutions for the base cases: \semisortequal{} and \semisortless{}.

\myparagraph{\semisortequal{}. }
In the base case, \semisortequal{} uses a sequential hash table with chaining.
We first build a hash table of size $(1+\epsilon)n'$ for some constant $\epsilon>0$.
Then, we iterate over all keys and insert each key to the hash table with separate chaining.
Finally, all \record{s} are packed to the output by looping over the hash table in order.
Chaining allows the algorithm to maintain the order of the original input for \record{s} with the same key,
and thus our algorithm is stable.
Since each base case is small, the hash table can be maintained locally by each thread.

\myparagraph{\semisortless{}.} In the base case, \semisortless{} uses a standard comparison sort.
By using a stable comparison sort, we can also guarantee the stableness of our \semisort{}.

\subsection{In-place Optimization}
\label{sec:inplace}
Before the recursive call in \cref{alg:semisort}, we copy the temporary array $T$ back to $A$ (\cref{line:copy}).
This copy accesses the whole arrays $A$ and $T$, which is expensive in practice.
We note that we can save this copying by swapping the two arrays $A$ and $T$ in the recursive call.
Namely, we skip \cref{line:copy} and in \cref{line:recursion} we sort the light buckets in $T$, and use the corresponding part in $A$ as the other array to take the output.
For the base cases and the heavy buckets, if they happen to reside in $T$, we copy them back to $A$.
By doing this, we avoid the copying in \cref{line:copy}.
This reuses the auxiliary array $T$ and also avoids allocating new memory in every recursive level.
Since in most cases the recursion will reach the base case in two levels, the entire algorithm copies the data twice per \record{}, first from $A$ to $T$, and then from $T$ back to $A$.

Here we use ``in-place'' to indicate that the input and output of semisort are in the same array.
Our algorithm still uses $O(n)$ extra space. We will discuss how to reduce space usage in \cref{sec:conclusion}.

\subsection{Supporting Histogram and Collect-Reduce}
\label{sec:histogram}

Using our semisort algorithm, the histogram and \collectreduce{}
primitives can be supported with minor modifications.
%
Here we will elaborate on \collectreduce{} since \histogram{} can be
considered as \collectreduce{} with values always equal to 1 for all records.

We still use the \sampling{} step to determine the heavy keys.
Then in the \counting{} step, it is unnecessary to distribute the heavy keys to their corresponding \bucket{s}.
Instead, we first directly compute the reduced values (or counts for \histogram{}) for
the heavy records in each \block{} (all the \block{s} can be processed in
parallel), and then reduce the results of all \block{s}.
%
In the base-case of the \refining{} step, we use the version based
on hash tables.  When any duplication is identified, we directly combine their
values instead of chaining.
Since the algorithm is stable, it works on any associative reduce
functions (in particular, there is no need to be commutative).

Generally speaking, \histogram and \collectreduce can be significantly
faster than semisort when there are many heavy duplicate keys, as we do not need to distribute the heavy records and only need to distribute the ``locally reduced value'' for each heavy key in each \block.
When no or few duplicate keys are in the input, \histogram and \collectreduce can perform slightly slower than semisort.
This is because they perform almost identical computations as semisort to reorder \record{s}, but need an extra step to pack the keys and reduced values
into the output.

\subsection{Analysis and Parameter Choosing}\label{sec:para}

Our new semisort algorithm has three parameters: $l$ (subarray size), $n_L$ (light bucket number), and $\alpha$ (base case size).
Other parameters (e.g., the number of heavy \bucket{s} $n_H$) are set accordingly.
The values of $l$ and $\nlight$ are fixed for \emph{all levels of recursions}.
To ensure the space usage is $O(n)$,
we will assume $\nlight\le \B$ since the sizes of matrices $\countm$ and $\presum$ has size $\Theta(\nlight\cdot n/\B)$.
We also assume the sample set size $|S|=\nlight\log n=O(n)$.
In the following, we will use $n$ as the \emph{original problem size}, and use $n'$ as
the current size of the recursion.


We will analyze the cost bounds and show that our semisort algorithm is efficient under \emph{reasonable assumptions} of modern multi-core architecture.
Then we will show how to select the parameters in practice for the best practical performance.

\myparagraph{Theoretical Analysis}. We start with analyzing the number of recursion levels in our algorithm.

\begin{lemma}
\label{lem:levels}
The number of recursion levels is $O(\log_{n_L}(n/\basecase))$ \whp{} for both \semisortequal{} and \semisortless{}.
\end{lemma}
\begin{proof}
  From the same analysis from \gssb{}~\cite{gu2015top}, the number of records in each light bucket is $O(n/n_L)$ \whp.
  Therefore, the light bucket size shrinks by a factor of $\Theta(n_L)$ \whp{} in each level of recursion,
  and the number of recursive levels is $O(\log_{n_L}(n/\basecase))$ \whp{}.
\end{proof}

For simplicity in stating the bounds, we use $\boldsymbol{r}=O(\log_{n_L}(n/\basecase))$ to denote the \defn{number of recursion levels}.
We start with the work of the algorithms and present the result in \cref{thm:work}.

\begin{theorem}
  \label{thm:work}
  The work of \semisortequal{} is $O(rn)$ \whp{}.
  The work of \semisortless{} is $O(rn+n\log \alpha)$ \whp{}.
\end{theorem}
\begin{proof}
We first show the work analysis for \semisortequal{}.
We start with considering the top level of recursion.
As assumed above, the number of samples is $O(n_L\log n)=O(n)$,
and thus the \sampling{} step has $O(n)$ work.
In the \counting{} step, it takes $O(1)$ work per \record{} to find the \bucket{} it belongs to.
As mentioned above, we assume $\nlight\le \B$ so that the \countmatrix $\countm$ and prefix array $\presum$ have sizes $O(n)$, and computing prefix sum also has $O(n)$ work.
The step to distribute the \record{s} to array $T$ (lines \ref{line:distribute-start}--\ref{line:distribute-end}) is also $O(n)$ since each \record{} is processed once.
For each recursion level, this argument is still true, and the work of all the subproblems in one level adds up to $O(n)$.
Assuming $r$ recursion levels, the work before entering the base cases is $O(n)$ for both \semisortequal{} and \semisortless{}.
For \semisortequal{}, the work of each base case is $O(n')$, which gives $O(n)$ total work for all base cases.
For \semisortless{}, the work of each base case is $O(n'\log n')$, where $n'$ can be at most $\alpha$.
Therefore the total base-case work is $O(n\log \alpha)$ for \semisortless{}.
Combining the results gives the bounds in \cref{thm:work}.
\end{proof}

Although \semisortless{} has a higher work, the overhead is caused by the comparison sort in base cases.
However, the base cases fit in cache and are highly-optimized.
In the experiments \semisortless{} shows as good performance as \semisortequal{} in most cases.

We then analyze the span of \semisortequal and \semisortless{}, and show that they are highly parallel.

\begin{theorem}
\label{thm:span}
  The span of \semisortequal{} is $O((\B+\nlight\log n)r+\alpha)$ \whp{}.
  The span of \semisortless{} is $O((\B+\nlight\log n)r+\log n)$ \whp{}.
\end{theorem}
\begin{proof}
The \sampling{} step is executed sequentially with $O(\nlight\log n)$ span.
We note that this step can be easily parallelized~\cite{gu2015top}, but our implementation still performs it sequentially,
since it is cheap anyway.
For the distributing step, we have two sequential for-loops (\cref{line:seq-loop-1,line:seq-loop-2}), leading to $O(\B)$ span.
Computing the prefix sum ($\presum$ from $\countm$) has $O(\log n)$ span. 
In total, the span of one recursive level is $O(\B+\nlight\log n)$.
Hence, considering $r$ recursive levels, both algorithms have $O((\B+\nlight\log n)r)$ span before the base cases.
\semisortequal{} uses sequential hash tables in base cases, which leads to $O(\alpha)$ span.
\semisortless{} uses a comparison sort in base case, which can achieve $O(\log n)$ span \whp{} in theory~\cite{blelloch2020optimal} (our implementation coarsens the base case by using a sequential sort, since the base case size is small).
Combining the results above gives the bounds in \cref{thm:span}.
\end{proof}

Considering both work and span, the parallelism (defined by the ratio between work and span) for both algorithms is roughly $\Theta(n/\B)$ (in practice we choose $\B$ much larger than $n_L$ and $\alpha$).
Given the number of processors $P$ in a machine, our semisort algorithm achieves sufficient parallelism if we can set $n/\B=\Omega(P)$.

We analyze the I/O bound of the algorithms with our choices of parameters to make the bound optimal ($O(n/B)$).
We make the assumption that $M/B=\Omega(n^{1/2})$ (recall that $M$ and $B$ are cache size and cacheline size, respectively).
For reasonable values of $n\le 10^{12}$, this assumption is true for
both commodity machines (e.g., laptops) as well as more powerful servers.
We present our results in \cref{thm:io}.

\begin{theorem}
\label{thm:io}
  Assume $M/B=\Omega(n^{1/2})$, using parameters $n_L=\Theta(n^{1/4})$, $\alpha=\Theta(n^{1/2})$, and $l=\Theta(n^{3/4})$, both \semisortequal{} and \semisortless{} have I/O cost of $O(n/B)$ \whp{}.
\end{theorem}
\begin{proof}
Given the parameters in the theorem, the number of recursive levels is $r=O(1)$ \whp{}.
Therefore, we only analyze the top-level recursion.
Since the sizes of $\countm$ and $\presum$ are $O(\nlight\cdot (n/\B))=O(\sqrt{n})=O(M)$,
all memory accesses to arrays $\countm$ and $\presum$ fully fit into
cache except for the first access.
When $M/B=\Omega(n^{1/2})$ and $\alpha=\Theta(n^{1/2})$, we can choose $\alpha$ to fit the base cases in cache,
such that the base cases can be solved without using additional main memory access after loading the data to the cache.
The only cache misses are when accessing the input array $A$ and the buckets $T$.
The accesses to $A$ are all serial accesses.
For $T$, we are writing serially from ${(n_H+n_L)\cdot (n/l)}$ pointers as stored in the $\presum$ matrix.
Even when all the pointers are non-consecutive, only $(n_H+n_L)\cdot (n/l)=O(n^{1/2})$ cachelines are active at any time,
and they all fit in cache.
For every pointer, there is one cache miss every $B$ accesses to the array $T$.
Therefore, the total I/O cost to generate $T$ is $O(n/B)$.
Note that this analysis is true for both the root level (when input size is $n$), as well as the recursive levels (the total sizes of $\countm$ and $\presum$ for all subproblems in the same recursive level are still $(n_H+n_L)\cdot n/l=O(n^{1/2})$).
In summary, both \semisortequal and \semisortless have I/O cost $O(n/B)$ \whp{} assuming $M/B=\Omega(n^{1/2})$, which improves the $O(n)$ I/O bound of \gssb{} by a factor of $O(B)$.
The bound is optimal, since loading the input needs $\Omega(n/B)$ I/Os.
\end{proof}

Since I/O-efficiency is one of our main design goals, we use the parameters in \cref{thm:io} to present the work and span bounds below.

\begin{theorem}
  \label{thm:workspan}
  Assume $M/B=\Omega(n^{1/2})$, and parameters $n_L=\Theta(n^{1/4})$, $\alpha=\Theta(n^{1/2})$, and $l=\Theta(n^{3/4})$.
  \semisortequal{} has $O(n)$ work, $O(n^{3/4})$ span, and $O(n/B)$ I/O cost.
  \semisortless{} has $O(n\log n)$ work, $O(n^{3/4})$ span, and $O(n/B)$ I/O cost.
  All bounds are \whp{} in $n$.
\end{theorem}

\hide{
\yihan{I cannot understand the ``for ease of understanding'' paragraph...}
For the ease of understanding, we use one set of parameters to run though the analysis of work, span, and I/O.
The work and span bounds hold when $n_L=n_H=n^{\epsilon}$ for $0<\epsilon<1$, and $l\ge n_L$ (the span bound is parameterized).
The I/O bound is $O(n/B)$ when $l/(n_L+n_H)\ge \sqrt{n}$, and $n_L$, $n_H$, and $\alpha$ are $n^{\epsilon}$ for $0<\epsilon<1/2$ (can be different values).
}

\myparagraph{Parameters in our Implementations.}
The performance of our semisort algorithm is reasonably consistent for a large parameter range.
The best parameters of each input instance can be different, decided by input size, heavy record ratio, etc.
In our implementation and all experiments, we pick $n_L=2^{10}$, $l=n/5000$ (at most 5000 subarrays in all subproblems in one recursive level), and $\alpha=2^{14}$.
These numbers satisfy the conditions in the theoretical analysis in \cref{thm:workspan} when $n=10^8$ to $10^9$.
We set the number of samples $|S|=500\log n$, so we can have at
most $n_H=500$ heavy keys.
We set up these parameters to ensure that the matrices $C$ and $\presum$ and the base cases are small enough to fit in the last-level cache for modern multicore machines.

\hide{
\subsection{Implementation Details}

We note that although our algorithms are conceptually simple and similar to
some existing algorithms in some steps, our selection of parameters and careful
application of I/O-efficient optimization techniques also contributes to the
practicality of our algorithms. We briefly present an important optimization in
this section.
\xiaojun{We discuss parameters somethere else.}

One useful optimization is to avoid redundant data movements, since the
performance of sorting and semisorting algorithms are usually bottlenecked by
data movements.
This is the reason that many in-place algorithms can naturally achieve relatively good performance.
If the output elements are written to the input array, we say it is \defn{in-place}.
We note that our definition of \emph{in-place} is different from some other
papers, and instead only requires the output to be in the same array as input.
Otherwise, if they are written to a new array, we say it is \defn{out-of-place}.
To implement an in-place semisort algorithm efficiently, we use a temporary array in the \counting step,
such that the elements can be placed into this array without messing up the elements in the input array.
The copy step is followed by moving the elements from the temporary array to the output array.
To minimize data movements, for light buckets, we can use the temporary array as the input, and perform an out-of-place sort in the next level of recursion.
This optimization can keep the algorithms from moving the elements back and
forth in the temporary array and the output array and save an extra round of
copies.
We note that this technique only works for semisort and does not work for collect-reduce and histogram,
as the input/output types may not be the same in these two applications.
}

\section{Comparisons with Existing Algorithms}\label{sec:discussion}

\subsection{Improvements over \BGSS{}}\label{sec:discussion-gssb}

In this section, we compare and discuss the improvements of our algorithm(s)
over the existing semisort algorithm \BGSS{}.

\myparagraph{Flexible Interface}.
Recall that \BGSS{} requires hashed keys (integers) as input, which needs a pre- and post-processing to resolve collisions.
Our algorithm supports \emph{arbitrary key types} $K$ with $=_K$ or $<_K$, with a \userhash{}.
For integer keys, we provide the option to use the identity function, resulting in \defn{\semisortequali} and
\defn{\semisortlessi}, which can be much faster in many cases, although we note that
these versions do not admit as good theoretical bounds.
Our interface also supports histogram and collect-reduce with minor changes.

\myparagraph{Low Space Usage}.
In the \counting{} step, we compute the exact counts for the buckets, so when
distributing the keys, the total size of the buckets is $n$, instead of
$\Theta(n)$ as in \gssb{} (their buckets need to have a load factor smaller than 1 because of random scatter).
Other than space overhead, \gssb{} also needs a costly packing step.
%

\myparagraph{I/O-Efficiency}.
Our algorithm also uses several techniques to enable a better memory access
pattern.  We pick a small number of \bucket{s} ($n_L=2^{10}$), as opposed to
$O(n/\log^2 n)$ of them in \gssb, such that the \countmatrix $C$ and its prefix sum
$\presum$ in our algorithms fit in cache (recall that we access them in column-major).  As such,
the \counting{} step incurs no random accesses to the main memory.


\hide{
Another reason that we can achieve better I/O-efficiency is the use of recursions, which can
\emph{identify more heavy keys}. 
For instance, if our algorithms recurse for two levels, we can identify at most $n_H^2\approx 2^{18}$ heavy keys.
For example, with levels of recursion, we can identify about $n_L^2=2^{20}$ heavy keys, more than $2^{16}$ in \gssb{}.
\xiaojun{What is $2^{16}$?}
If we wanted to find $2^{20}$ heavy keys in a non-recursive manner, the \countmatrix $C$ no longer fits into the cache and will cause excessive random accesses.
}
\myparagraph{Stability and Determinism}.
Due to avoiding using parallel hash tables, our semisort algorithms (both \semisortequal and \semisortless) are stable and race-free.
\gssb{} is not race-free (due to using parallel hash tables), and is unstable (heavy keys are in random order), and thus cannot support non-commutative operations in collect-reduce.

\subsection{Relationship to Sample Sort and Integer Sort}\label{sec:discussion-sort}

Many ideas in our semisort algorithm are closely related to ideas in sorting algorithms, as we will discuss in this section.

\myparagraph{Samplesort}.
Samplesort is the general idea of using multiple pivots in quicksort; clearly this algorithm can be used to solve the \semisortless{} problem.
The algorithm selects $p$ pivots, uses them to partition the input into $p+1$ buckets, 
and sorts all of the buckets in parallel. We refer the audience to~\cite{axtmann2022engineering} for a detailed literature review on samplesort.
We compare to the state-of-the-art samplesorts from ParlayLib~\cite{blelloch2020parlaylib} and IPS$^4$o~\cite{axtmann2022engineering} in our experiments.

Similar to samplesort, our algorithm also partitions the input into
buckets and processes them in parallel.
However, samplesort is a comparison sort that requires the $<_k$ operation, and
has an $\Omega(n\log n)$ lower bound in work, whereas our \semisortequal{}
algorithm only requires $O(n)$ work.
%
The ParlayLib samplesort~\cite{blelloch2020parlaylib} uses one level of partition. 
IPS$^4$o~\cite{axtmann2022engineering} (preliminary version as~\cite{sanders2004super}) also uses a small number of samples and sort recursively.
They use an implicit search tree (breadth-first traversing the tree that stores the sorted pivots) to find the bucket for each record, which is not required in our approach. 
They also use a smart approach for the distribution step, and we discuss this in \cref{sec:conclusion}.
 in a variety of places.
%

\myparagraph{Integer sort}.
Integer sorting can be used to semisort integer keys for \semisortequal{}, or to semisort the hash value of any key types with an extra step to resolve collisions.
Unlike the sequential radix sort that starts from the least-significant bits, all parallel integer sort algorithms are top-down and first look at the most-significant bits.
We refer to~\cite{obeya2019theoretically} for a detailed literature review for parallel integer sort.
We compare to the state-of-the-art integer sorts from ParlayLib~\cite{blelloch2020parlaylib}, RegionsSort~\cite{obeya2019theoretically}, and IPS$^4$Ra~\cite{axtmann2022engineering} in our experiments.

The major advantage of our semisort algorithm over integer sorting is that our algorithm can identify heavy keys.
Consider a heavy key $x$ and a light key $x+1$.
Our algorithm will put $x$ in a separate heavy \bucket{}, and only deal with $x+1$ in a light bucket in the last step.
For existing integer
sorts~\cite{blelloch2020parlaylib,obeya2019theoretically,axtmann2022engineering},
both keys are likely kept in the same bucket for all levels and
separated only in the last round, which can result in significant
wasted work and load imbalance.

\hide{
\smallskip
We believe that our new semisort algorithm and implementation also provides new
insights that can be used in implementing more efficient samplesort and integer
sort algorithms, which we leave as future work.
}

\hide{
\begin{algorithm}
\caption{The Semisort Framework (\SMS{})}\label{algo:framework}

\textbf{\sampling{}.} Sample and determine the heavy keys.
Heavy keys use an individual bucket; multiple light keys share a \bucket{}.\\

\textbf{\counting{}.} Split the input into \block{s} and process each \block{} in parallel. Within each \block{}, count the \record{s} falling into each \bucket{}. \\

\textbf{\placing{}.} Assign appropriate memory for each \bucket{}, Then in parallel, process each \block{} to place each element into its corresponding \bucket{}.\\

\textbf{\refining{}.} Locally \semisort{} the light \bucket{s} to gather identical keys in the same \bucket{}.

\end{algorithm}

\begin{itemize}
\item multiple levels allows us to identify more heavy keys; heavy keys are good for the algorithm: better parallelism, better load balancing, anything more?
\item no assumption that the inputs are hashed keys; our framework is more general and supports arbitrary records and keys equipped with lt or eq,
\item (related to generality) supporting both \semieq{} and \semile{},
\item (related to generality) supporting generalizations of semisort (histogram, collect-reduce),
\item significantly lower space usage,
\item better cache-performance (could quantify this experimentally by measuring L3 misses, if time).
\item Our algorithm is (can be) stable, but \gssb{} is not stable (heavy keys are put in random order).  We should also mention this early on.
\end{itemize}

\laxman{We have presented GSSB in Section 3, and there is a decent amount of overlap between that description and our description here. I believe an unfamiliar reader will be confused as to what is new here. Shall we try to very explicitly comment on this in a sub-section on \emph{what is new in our framework} from GSSB? Things we have discussed include:
\begin{itemize}
\item multiple levels allows us to identify more heavy keys; heavy keys are good for the algorithm
\item no assumption that the inputs are hashed keys; our framework is more general and supports arbitrary records and keys equipped with lt or eq,
\item (related to generality) supporting both \semieq{} and \semile{},
\item (related to generality) supporting generalizations of semisort (histogram, collect-reduce),
\item significantly lower space usage,
\item better cache-performance (could quantify this experimentally by measuring L3 misses, if time).
\end{itemize}
I guess some of these points are experimental, but if we clarify how these points became possible due to the framework/algorithm design, they would fit in this discussion. Another place we could have this discussion is after Section 5.}

\yihan{We need to update the code and add some references of line numbers to the algo description. Also add some illustration figures. }

}

%% file: algo-semisort-new.tex
\definecolor{commentgreen}{RGB}{0,128,0}
\newcommand\mycommfont[1]{\textit{\textcolor{commentgreen}{#1}}}
\SetCommentSty{mycommfont}
\renewcommand{\LinesNumbered}{%
  \setboolean{algocf@linesnumbered}{true}%
  \renewcommand{\algocf@linesnumbered}{\everypar={\nl}}}%
\newcommand{\nosemic}{\renewcommand{\@endalgocfline}{\relax}}
\newcommand{\dosemic}{\renewcommand{\@endalgocfline}{\algocf@endline}}
\newcommand{\pushline}{\Indp}
\newcommand{\popline}{\Indm\dosemic}
\let\oldnl\nl
\newcommand{\nonl}{\renewcommand{\nl}{\let\nl\oldnl}}
\makeatother
\newcommand{\stepcomment}[1]{\vspace{.03in}\nonl \popline\textbf{\textcolor{commentgreen}{\enspace\,\textit{\underline{#1:}}}}\\\pushline}
\begin{algorithm}[t]
\small
    \caption{The Semisort Algorithm\label{alg:semisort}}
    \KwIn{The input array $A$, a \userhash{} $h$, and a comparison function $\comp$ ($=$ or $<$). The original (top-level) input size is $n$,
    and the current subproblem size is $n'$.}
    \KwOut{The semisort result in $A$ (in-place)}
    \SetKwInOut{Parameters}{Parameters}
    \SetKwProg{myfunc}{Function}{}{}
    \SetKwFor{parForEach}{parallel\_for\_each}{do}{endfor}
    \SetKwFor{parFor}{parallel\_for}{do}{endfor}
\Parameters{
$\nlight=2^b$: number of light buckets.\\
$\basecase$: base case threshold.\\
$\B$: subarray size.\\
}
    \lIf(\tcp*[f]{Base cases}) {$|A| < \basecase$} {
        \Return{BaseCase$(A,h,\comp)$ } \DontPrintSemicolon \label{line:basecase}
    }
\stepcomment{\sampling}
    $S \gets \nlight\log n'$ sampled keys from $A$ \label{line:sample}\\
    Count the occurrences of each key in $S$ \label{line:count}\\
    Initialize the heavy table $H$ \\
    $id\gets \nlight{}$\\
    \tcp{This for-loop can also be performed in parallel theoretically}
    \For {each distinct key $k \in S$ } {
        \If {the occurrences of $k$ in $S$ is at least $\log n$} {
            $H.$insert$(k,id)$ \label{line:insert}\tcp*[f]{Assign bucket id $i$ to heavy key $k$}\\
            $id\gets id+1$
        }
    }
    $\nheavy \gets$ number of distinct keys in $H$ \label{line:step1:end}\\
    \stepcomment{\counting}

    Initializing matrix $C[][]$ with size $(\nlight+\nheavy)\times(n'/\B)$\\
    \parFor(\tcp*[f]{\mbox{For each \block{}}}){$i: 0\le i < n'/\B$\label{line:compute-c}}{
      \For{$j: i\cdot \B \le j < (i+1)\cdot\B$\label{line:seq-loop-1}} {
        $id\gets \textsc{GetBucketId}(\key(A[j]),H,h,\nlight)$\\
        \tcp*[h]{\mbox{$C[i][id]$:~\#\record{s} falling into bucket $id$ in \block{} $i$}}
        $C[i][id]\gets C[i][id]+1$\label{line:cplus}
      }
    }
    Initialize $T$ of size $n'$\\
    \tcp*[h]{\mbox{$\presum[i][j]$:~offset in $T$ for record in \block{} $i$ going to \bucket{} $j$}\label{line:offset}}
    Compute $\presum[i][j]\gets \sum_{j'<j\text{~or~}(j'=j, i'<i)}C[i'][j']$\label{line:prefix}\\
    \parFor{$i: 0\le i\le \nlight+\nheavy$}{$\offsets[i]\gets \presum[i][0]$\label{line:step2:block}}
    \parFor(\tcp*[f]{\mbox{For each \block{}}}){$i: 0\le i < n'/\B$\label{line:distribute-start}}{
      \For{$j: i\cdot \B \le j < (i+1)\cdot\B$\label{line:seq-loop-2}} {
        $id\gets \textsc{GetBucketId}(\key(A[j]),H,h,\nlight)$\\
        $T[\presum[i][id]]\gets A[j]$\\
        $\presum[i][id]\gets \presum[i][id]+1$\label{line:distribute-end}\\
      }
    }
    $A\gets T$\tcp*[f]{\mbox{Avoided in implementation, see \cref{sec:inplace}}}\label{line:copy}\\
    \stepcomment{\refining}
    \parFor(\tcp*[f]{Only for light buckets}) {$i: 0\le i< \nlight$} {
        \mf{Semisort}$(A[\offsets[i].. \offsets[i+1]], h, \comp)$ \label{line:recursion}
    }
    \Return{$A$}\\
\vspace{.05in}
    \myfunc{\upshape \mf{GetBucketId}$(k, H, h, n_L)$\label{line:bucketid}} {
        \lIf {$k$ is found in $H$} {
            \Return{the heavy id of $k$ in $H$} \DontPrintSemicolon
        } \lElse(\tcp*[f]{$h(\cdot)$ is the hash function}) {
            \Return{$h(k) \mod n_L$} \DontPrintSemicolon
        }
    }
    \end{algorithm}

%% file: figures/fig-semisort.tex
\begin{figure*}
  \centering
  \includegraphics[width=2\columnwidth]{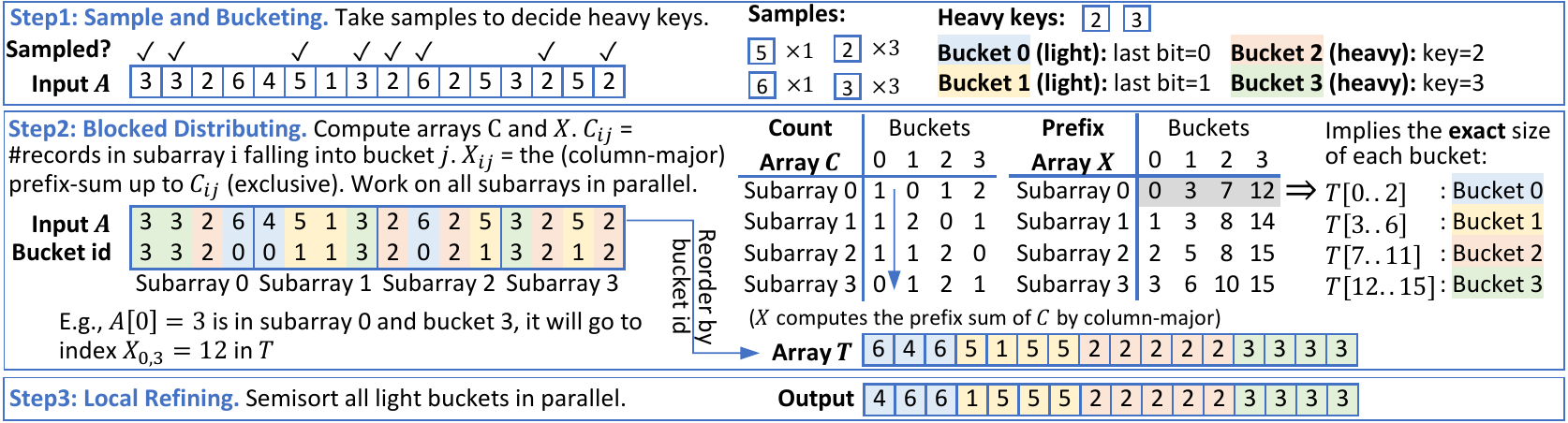}
  \caption{\small \textbf{Our algorithm with a running example.} We consider an input with $n=16$ records with keys given.  $|S|=8$ samples are taken, and keys with more than $2$ samples are heavy keys.
  $\nlight=\nheavy=2$ in this example.
  We have $\B=4$ subarrays each with 4 records.  
  We compute the \countmatrix $C$ and prefix array $\presum$ as shown, and records can be distributed accordingly.  Finally the local refining step recursively solves the 2 light buckets.
  }\label{fig:semisort}
  \vspace{-2em}
\end{figure*} 

%% file: exp.tex
\newcommand{\impname}[1]{\textsf{#1}}
\newcommand{\ipso}{\impname{IPS$^4$o}\xspace}
\newcommand{\gssbimp}{\impname{GSSB}\xspace}
\newcommand{\region}{\impname{RS}\xspace}
\newcommand{\plss}{\impname{PLSS}\xspace}
\newcommand{\plcr}{\impname{PLCR}\xspace}
\newcommand{\plis}{\impname{PLIS}\xspace}
\newcommand{\ipsra}{\impname{IPS$^2$Ra}\xspace}
\newcommand{\competitive}[1]{within #1\%}
\newcommand{\uniform}[1]{\emph{uniform-#1}}
\newcommand{\exponential}[1]{\emph{exponential-#1}}
\newcommand{\zipfian}[1]{\emph{Zipfian-#1}}

\section{Experiments}\label{sec:exp}

\hide{
\yihan{Some figures need to be polished:

1, increase the font and marks of almost all figures

2, double check titles of the axis.

3, for some experiments we only show 1 or two distributions (among uniform, zipfian and exponential). I'm not sure if we are OK with a subset of them or we plan to add all of them.

4, for all figures, we should change the legends: a) use \semisortlessi{} and \semisortequali{} when it is integer-only (instead of ``no\_hashing''),
b) use PLSS for parlay::samplesort, PLIS for parlay::integersort, and RS for regionsort. It's better using both names, e.g., PLSS (parlay::samplesort),
c) use \semisortequal{} and \semisortless{} instead of semisort\_equal and semisort\_less.

5, in \cref{fig:collect-reduce}, maybe add something to indicate which side has more heavy keys. Also, use 0.6, 1 and 2 on x-axis, instead of $6\times 10^{-1}$, $10^0$ and $2\times 10^0$.

6, all distribution statistics are missing.

7, we probably need to reorder them to put 3-4 figures in 2 columns.

8, add two columns for distinct keys and max occurrence in the table. Reorder all rows inside each distribution, e.g., make all of them to go from more even to more skew. Maybe add another row to indicate if they are stable or not.  We also need to update the heatmap accordingly.

9, we can remove the global hash table version in collect-reduce from the figure. It doesn't have good performance anyway.

10, The results in figure 5 look incorrect. I'm not sure for the other figures. We should double-check. Also, many of the figures only show one version of \semisortequal{} and \semisortless{}. I'm not sure if they are with hashing or not.

11, we also need to prepare all figures and tables in the appendix.

12), the figure for ngram is not in the paper.
}
}

\myparagraph{Experimental Setup.}
We run our experiments on a 96-core machine (with two-way hyper-threading) with 4 $\times$ 2.1 GHz Intel Xeon Gold 6252 CPUs processors (with 36MB L3 cache) and 1.5TB of main memory.
We implement our algorithms in C++ using ParlayLib~\cite{blelloch2020parlaylib}
for fork-join parallelism and some parallel primitives.
We compile our code using \texttt{clang} version 14.0.6 with \texttt{-O3} flag.
We always use \texttt{numactl -i all} to interleave the memory on all CPUs except for sequential tests. 
We run each test four times and report the median of the last three runs.
All running times are given in \textbf{seconds}.

\input{tables/baseline.tex}

\input{tables/distribution.tex}
\myparagraph{Baseline Algorithms.}
We compare our algorithms to the state-of-the-art comparison and integer sorting algorithms and collect-reduce algorithms.
We provide the list of the baseline algorithms we compare our algorithms against in \cref{tab:baseline}.
For fairness and consistency, we require the output to be written to the input
array (i.e., in-place).
We note that this is beneficial for \plss{}, \ipso, \ipsra, and \region as they are originally designed for the in-place setting.
%
%
Some of the baselines only work for integer types (\defn{integer-only}),
including \plis{}, \gssbimp{}, \region{}, and \ipsra{}.
\ipso and \plss work on
any input types (\defn{any-type}).
For the any-type algorithms, \semisortless{}, \plss{} and \ipso{} require the
less-than test $<_K$, while our \semisortequal{} only needs the equality-test~$=_K$.
%
\gssbimp{} assumes the input keys are already hashed and does not resolve
collisions, so we also categorize it as integer-only.
Among all implementations, all our algorithms and \plis{} are \emph{stable}.
This also means that they can be applied to \collectreduce{} with arbitrary (associative) reduce operations,
while the others also require the reduce operation to be commutative.
We note that there are two versions of samplesort in ParlayLib.
The stable one is slower and the unstable one is faster.
Our experiments use the unstable but faster version.
When comparing the \emph{average} performance, we always use the \defn{geometric mean}.


\myparagraph{Our Algorithms.}
We use the two versions of our algorithm \semisortless{} and \semisortequal{} that work on \defn{any-type}.
In tables and figures, we also use ``Ours$_=$'' and ``Ours$_<$'' to refer to them, and use ``Ours$_\oplus$'' to refer to our collect-reduce implementation.
When comparing with the integer-only implementations, we use simplified
versions without hashing (see
\cref{sec:discussion-gssb}), and call them \semisortequali{} and \semisortlessi{} (or ``Ours-i$_=$'' and ``Ours-i$_<$''),
where the hash function is an identity function.
The choices of parameters in our algorithms are in \cref{sec:para}.

%
%

\myparagraph{Input Distributions.}
We use three distributions for evaluating our algorithms: uniform$(\mu)$, exponential$(\lambda)$, and Zipfian$(s)$.
If not specified, the default setting is $n=10^9$ with 64-bit keys and 64-bit values.
We also include tests of our algorithm on varying input sizes and key lengths (\cref{fig:input-size-scaling-zipfian,fig:scaling-keylengths-zipfian}).
For uniform distribution, we test $\mu=10^1, 10^3, 10^5, 10^7, 10^9$.
For exponential distribution, we test $\lambda=1\times 10^{-5},2\times 10^{-5},5\times 10^{-5},7\times 10^{-5},1\times 10^{-4}$.
For Zipfian distribution, we test $s=0.6,0.8,1,1.2,1.5$.
We use \emph{distribution-param} to denote the input distribution with parameter \emph{param} (e.g., \uniform{$10^9$}).
We show relevant statistics of the inputs along with our results in \cref{tab:distribution}.
We present the number of distinct keys, the maximum frequency, and the ratio of keys with more than $500\log n$ occurrences, which is noted as ``Heavy Freq.'' in \cref{tab:distribution}.
They are measured for each distribution to indicate skewness of the data.
For the synthetic data, we always set the value type the same as the key type.
For most of the tests, we provide figures on one representative distribution, and provide more results in the appendix.

\input{figures/scale.tex}

\input{figures/input-type}

\subsection{Overall Performance}

We present the running time of all tested implementations with $n=10^9$ 64-bit keys with different distributions in \cref{tab:distribution},
and a heatmap (normalizing all running times to the fastest on each test) in \cref{fig:heatmap}.
On all but one test, our algorithms are always the \defn{best two} implementations.
Among any-type algorithms, our \semisortequal{} and \semisortless{} are 1.03--2.47$\times$ and 1.04--2.53$\times$ faster (respectively) over the best of the other algorithms.
For integer-only algorithms, our \semisortequali{} is 1.09--2.78$\times$ faster than the other algorithms.
Our \semisortlessi{} is about 20\% slower than \plis{} in one test, and is up to 2.82$\times$ faster than all baselines on all other tests.

Overall, our algorithms are always
faster than the baseline algorithms using geometric mean.
Note that some of the baselines
are competitive on some individual tests, such as
\ipso{} on \uniform{$10^3$} and \uniform{$10^9$},
\plis{} on \uniform{$10^9$} and \zipfian{$0.6$},
and \ipsra{} on \zipfian{$0.6$}.
However, their performance can be unstable over different distributions.
\ipso{} is relatively fast on uniform distributions
but performs worse on skewed distributions.
We also compute the geometric means in \cref{tab:distribution} and \cref{fig:heatmap}
to compare the performance on each distribution.
Based on these numbers, \semisortequal{} and \semisortless{} have very close
performance (within 5\%).  All the other algorithms are at least 30\% slower
than both of our implementations on average.
We also show relative performance for 32-bit and 128-bit keys in \cref{fig:heatmap-32,fig:heatmap-128}.
On average, our algorithms are consistently the fastest.
We note that not all comparisons are apple-to-apple comparisons.
\plss{} and \ipso{} work for general sorting which is
asymptotically harder than semisort.
\plis{}, \region{}, and \ipsra{} are for integer sorting, which is
also slightly different than semisorting.
Also, \plis{} and all our implementations are stable while others are not (see \cref{tab:baseline}).

Interestingly, the integer sort algorithms can be
slower than comparison sorts on 64-bit keys.
We tested on 32-bit and 128-bit keys and show the running time in \cref{fig:key-size-uniform-1e3,fig:key-size-uniform-1e7,fig:key-size-exponential-2e-5,fig:key-size-exponential-7e-5,fig:key-size-Zipfian-0.8,fig:key-size-Zipfian-1.2} in the appendix.
Unsurprisingly, integer sort algorithms are usually faster than comparison sort
algorithms on 32-bit keys, and get worse on 128-bit keys (\plis{} is the only integer sort in \cref{tab:baseline} that supports 128-bit keys).
On average, our algorithms are still the fastest on 32- and 128-bit keys, and the gap is smaller for 32-bit keys and larger for 128-bit keys.

One advantage of our algorithms is that they can identify
heavy keys and use little further work (no local refining needed) on them.
Thus, the running time of our algorithms decreases with more heavy keys (see \cref{tab:distribution}).
Many baseline algorithms also use optimizations on the heavy keys (e.g., \plss{}), and they show a similar trend.

\myparagraph{Parallel Scalability.}
We present the scalability curves using different number of threads in \cref{fig:speedup-zipfian} on
one representative distribution (\zipfian{$1.2$}, $n = 10^9$), and for other distributions in \cref{fig:scalability-uniform-1e3,fig:scalability-uniform-1e7,fig:scalability-exponential-2e-5,fig:scalability-exponential-7e-5,fig:scalability-Zipfian-0.8,fig:scalability-Zipfian-1.2} in the appendix.
All of our semisort algorithms, as well as \plss{}, generally achieve the top-tier (almost linear) speedup,
while other algorithms also scale well with increasing core counts.
The self-speedup of \semisortequal{} and \semisortless{} are 50--80$\times$,
The speedup numbers are slightly worse for \semisortequali{} and \semisortlessi (30--50$\times$ speedups),
as they save the work for the hashing step but can lead to unbalanced subproblem partitioning (light buckets).


\myparagraph{Input Size Scalability.}
We test all algorithms on input sizes from $10^7$ to $10^9$ on different distributions.
A representative one (\zipfian{}{$1.2$} is given in \cref{fig:input-size-scaling-zipfian}), and others are given in appendix (\cref{fig:input-size-uniform-1e3,fig:input-size-uniform-1e7,fig:input-size-exponential-2e-5,fig:input-size-exponential-7e-5,fig:input-size-Zipfian-0.8,fig:input-size-Zipfian-1.2}).
For very small test cases $n\le 2\times 10^7$, \plss{} is the fastest on certain tests.
However, in those cases, the running time is below 0.05s.
For $n\ge 5\times 10^7$, our algorithms are consistently faster than all baselines.
These results indicate that our algorithms perform well on reasonably small size and scale favorably well to large inputs.

%

\myparagraph{Varying Key Lengths.} In addition to 64-bit keys, we also tested 32-bit and 128-bit keys for $n=10^9$.
We always set the value to be the same type as the key.
Full results are given in the appendix (\cref{fig:key-size-uniform-1e3,fig:key-size-uniform-1e7,fig:key-size-exponential-2e-5,fig:key-size-exponential-7e-5,fig:key-size-Zipfian-0.8,fig:key-size-Zipfian-1.2}), and the running times on one representative distribution are shown in
\cref{fig:scaling-keylengths-zipfian} (\zipfian{$1.2$}).
Firstly, integer sort algorithms are sensitive to key lengths.
\region{} and \ipsra do not support 128-bit keys, and \plis's performance on 128-bit keys is usually the slowest based on our testing.
On 32-bit keys, integer sort algorithms can achieve much better (relative) performance than on 64-bit keys.
Also, integer sort algorithms generally perform poorly on highly-skewed data (see discussion in \cref{sec:discussion-sort}).
Other algorithms, including semisort (ours and \gssb) and comparison sort (\plss and \ipso), are less sensitive to key lengths.
Hence, the trends on 32- and 128-bit are similar to that on 64-bit.
Our new algorithms generally perform well since semisort is simpler than sorting, and we can apply special optimizations (e.g., for heavy keys).
In certain cases when not many special optimizations can be used (e.g., \uniform{$10^9$}), \plss and \ipso perform similarly to our algorithms.
\hide{
\ipso also performs well on 32-bit inputs.
It is faster than our implementations in certain test cases,
and performs consistently across different distributions.
On 128-bit keys, our implementations achieve significantly better overall performance.
Our algorithms perform well since 1) they do not depend on the key lengths (unlike integer sort), and 2) we make special optimizations for heavy keys.
On average, our implementations are 30\% faster than \ipso, 80\% faster than \plss{},
and at least 3$\times$ faster than other implementations.
}
\subsection{Collect-Reduce}

\input{tables/applications}

We test our \collectreduce{} algorithm  (\histogram is a special case for \collectreduce) and show the results on Zipfian distribution in \cref{fig:collect-reduce}.
The full results for other distributions are given in \cref{fig:cr-uni,fig:cr-exp,fig:cr-zip}.
Recall that our \collectreduce{} algorithm is similar to \semisortequal{},
but directly combines values for keys.
The values of the heavy keys are combined in the \counting{} step (no need to distribute),
and the values of the light keys are combined in the \refining{} step.
The only existing parallel implementation of \collectreduce{} that we know of is in ParlayLib~\cite{blelloch2020parlaylib} (\plcr{}), and we compare with it.
We also show the performance of \semisortequal{} as a baseline in \cref{fig:collect-reduce}.
The operator that we test for the reduce (on the values) is addition.
We use Zipfian distributions with varying parameters as it smoothly covers different amounts of skew in the input.
First, our \collectreduce{} is consistently faster than ParlayLib's implementation, and the gap is larger when the distribution is more skewed.
Furthermore, when heavy keys occur more, \collectreduce{} is significantly faster than \semisortequal{}.
This is because we reduce the values for each bucket in the \counting{} step, and then combine them without moving them.
However, when few heavy keys exist, \collectreduce{} incurs more work than \semisort{}, because some additional work is needed in the \refining{} step to pack the output since some keys are combined,
while in \semisort{} the input size equals to output size and no packing is needed.
In conclusion, when the input is more skewed (more heavy keys),
\collectreduce{} is faster than \semisortequal{}, and vice versa on more
evenly-distributed data (more light keys).

\subsection{Applications}
We integrate our algorithms into two real-world applications---graph transposing, where the input is edges,
and n-grams, where the input is strings---to test our algorithm in more realistic settings.
Unlike our previous experiments with synthetic distributions for performance study,
here we benchmark these applications on real-world
datasets and derive a more realistic understanding of semisorting performance
in practice.

\myparagraph{Graph transposing.}
Our first application is to transpose a directed graph $G=(V,E)$, i.e.,
to generate $G^\matrixtrans=(V,E^\intercal)$, where
$E^\matrixtrans=\{(u,v): (v,u)\in E\}$.
This is a widely used primitive in graph algorithms.  For example,
parallel algorithms for strongly connected components~\cite{blelloch2016parallelism,gbbs2021,ji2018ispan,slota2014bfs} require running
reachability searches both ``forwards'' and ``backwards''.
The backward reachability searches can be performed by running forward
reachability query on $G^\matrixtrans$.
%
In many existing graph libraries, the edges are stored in the
Compressed Sparse Row (CSR) format, where for each vertex $v$, the other
endpoints of edges from $v$ are stored contiguously.  Thus, transposing
the graph is exactly semisorting the CSR input using the other endpoint.  In
some existing parallel graph libraries such as Ligra~\cite{shun2013ligra} and
GBBS~\cite{gbbs2021}, \emph{stable} comparison sorts are used for
graph transposing to preserve the ordering of the first endpoint.


We evaluate transpose on four real-world directed graphs, soc-LiveJournal
(LJ)~\cite{backstrom2006group}, twitter (TW)~\cite{kwak2010twitter}, Cosmo50 (CM)~\cite{cosmo50,wang2021geograph},
and sd\_arc (SD)~\cite{webgraph}, where
the largest input has 2.04 billion directed edges.
For the social networks (LJ, TW) and web graph (SD), the degree distributions are more skewed.
For the $k$-NN graph CM, the degrees are more evenly-distributed.
We give more details about these datasets in~\cref{tab:graph_transposing}.
We use the initial CSR versions of these graphs and use our \semisortless{} and \semisortequal{} algorithms to transpose the graphs.
We compare with all the baseline algorithms and show the relative performance in \cref{tab:graph_transposing}.
On all the graphs, the keys (vertex id) are 32-bit.
Since the input data are integers, we use our integer version (identity hashing function).

Our \semisortequali{} is the fastest on three graphs (TW, CM, and SD), and is within 15\% slower than the fastest on the other graph (LJ).
Our \semisortlessi{} is competitive, and is within 20\% slower than the fastest on the other graphs.
\plis{} has relatively good performance on all graphs; it is the fastest on LJ (the smallest graph) and within 35\% on the others.
On the average performance across the four graphs,
\semisortequali{} is significantly better than the others (1.15--2.13$\times$ faster).
\semisortlessi{} and \plis{} have similar performance on average (within 1\%).
They are at least 25\% faster than other implementations.

\myparagraph{N-Gram.} Our second application is to process $n$-grams, where an
$n$-gram is a consecutive sequence of $n$ items from a given sequence (e.g., text
or speech).
We use the 2-gram and 3-gram datasets from Wikipedia~\cite{wikipediangram}, and clean the data by only
keeping alphabetical characters and converting them to lowercase.
Each $n$-gram record consists of $n$ consecutive words in the document.
We use the first $n-1$ words of a record as the key, and use the last word as the value.
We note that in our algorithms, we compute the hash values of the words on the fly.
Semisorting $n$-grams can be used to identify all possible words after a given
context, and to provide recommendations for text inputs, and to learn the pattern
of the input sequences.
Our results are shown in \cref{fig:ngrams}.
On both 2-gram and 3-gram, our \semisortequal{} is the fastest, while \semisortless{} (within 25\% slower) is competitive.
The average performance of \semisortequal{} is 15\% faster than \semisortless{},
24\% faster than \plss{}, and 2.3$\times$ faster than \ipso{}.

%

\hide{
including
the GSSB semisort~\cite{gu2015top} (\gssbimp{}),
the Regions sort~\cite{obeya2019theoretically} (\region{}),
the sample sort and integer sort from ParlayLib~\cite{blelloch2020parlaylib} (\plss{} and \plis{}),
the sample sort \ipso{}~\cite{axtmann2022engineering},
the radix sort \ipsra{}~\cite{axtmann2022engineering},
and the collect reduce from ParlayLib~\cite{blelloch2020parlaylib} (\plcr{}).
}

%% file: tables/baseline.tex

\begin{table}
    \centering\small
    \begin{tabular}{@{  }@{  }l@{  }c@{  }c@{  }@{  }c@{  }c@{  }@{  }l@{  }}
      \toprule
      \bf Name & \bf Stable & \bf Det.&\bf $K$ & \bf \comp{} & \bf Notes\\
      \midrule
      \textbf{Ours$_=$} &  Yes & Yes & Any& $=$ &Our \semisortequal{} algorithm\\
      \textbf{Ours$_<$} &  Yes & Yes & Any& $<$ &Our \semisortless{} algorithm\\
      \textbf{Ours-i$_{=}$} &  Yes & Yes & Int& $=$ &Our integer \semisortequal{} algorithm\\
      \textbf{Ours-i$_{<}$} &  Yes & Yes & Int& $<$ &Our integer \semisortless{} algorithm\\
      \textbf{Ours$_{\oplus}$} &  Yes & Yes & Any& $=$ &Our collect-reduce algorithm\\
      \textbf{PLSS} &  Y/N & Yes & Any & $<$ &ParlayLib sample sort~\cite{blelloch2020parlaylib}\\
      \textbf{PLIS} & Yes & Yes & Int& $<$ &ParlayLib integer sort~\cite{blelloch2020parlaylib}\\
      \textbf{IPS$^4$o} & No & No & Any&$<$ & IPS$^4$o sample sort~\cite{axtmann2022engineering}\\
      \textbf{IPS$^2$Ra} & No & No & Int&$<$ & IPS$^2$Ra integer sort~\cite{axtmann2022engineering}\\
      \textbf{GSSB} & No & No & Int&$=$ & GSSB semisort~\cite{gu2015top}\\
      \textbf{RS} & No & No & Int&$<$ & RegionsSort~\cite{obeya2019theoretically}\\
      \textbf{PLCR} & Yes & Yes & Any&$<$ & Collect-reduce from ParlayLib~\cite{blelloch2020parlaylib}\\
      \bottomrule
    \end{tabular}\vspace{-.15in}
    \caption{\small \textbf{Algorithms tested in our experiments.} ``Det.'' = determinism. ``$K$'' = key type. ``Any'' = any input key type. ``Int'' = only allows for integer keys.
    ``\comp'' = required comparison function.
    \plss{} has two implementations. We use the faster but unstable version in our experiments. 
    \label{tab:baseline}}\vspace{-.1in}
  \end{table} 

%% file: tables/distribution.tex
\begin{table*}[htbp]
  \centering
  \small


\begin{tabular}{rrrrr|rrrr|rrrrr@{}r}
  \multicolumn{1}{r|}{} & \multicolumn{1}{c}{\textbf{Para-}} & \multicolumn{1}{c}{\textbf{Dist.}} & \multicolumn{1}{c}{\textbf{Max}} & \multicolumn{1}{c|}{\textbf{Heavy}} & \multicolumn{4}{c|}{\textbf{Any Type}} & \multicolumn{6}{c}{\textbf{Integer Only}} \\
  \multicolumn{1}{r|}{} & \multicolumn{1}{c}{\textbf{meter}} & \multicolumn{1}{c}{\textbf{ Keys}} & \multicolumn{1}{c}{\textbf{Freq.}} & \multicolumn{1}{c|}{\textbf{Freq.}} & \multicolumn{1}{c}{\boldmath{}\textbf{Ours$_=$}\unboldmath{}} & \multicolumn{1}{@{ }c}{\boldmath{}\textbf{Ours$_<$}\unboldmath{}} & \multicolumn{1}{c}{\textbf{PLSS}} & \multicolumn{1}{@{  }c|}{\boldmath{}\textbf{IPS$^4$o}\unboldmath{}} & \multicolumn{1}{c}{\boldmath{}\textbf{Ours-i$_=$}\unboldmath{}} & \multicolumn{1}{@{}c}{\boldmath{}\textbf{Ours-i$_<$}\unboldmath{}} & \multicolumn{1}{c}{\textbf{PLIS}} & \multicolumn{1}{@{}c}{\textbf{GSSB}} & \multicolumn{1}{c@{ }}{\textbf{RS}} & \multicolumn{1}{@{ }c}{\boldmath{}\textbf{IPS$^2$Ra}\unboldmath{}} \\
  \midrule
  \multicolumn{1}{c|}{\multirow{6}[2]{*}{\begin{sideways}\textbf{Uniform}\end{sideways}}} & 10    & 10    & 100M  & 100\% & 0.675 & \underline{0.672} & 1.19  & 0.876 & 0.622 & \underline{0.606} & 1.81  & 2.76  & 1.26  & 3.84 \\
  \multicolumn{1}{c|}{} & $10^3$ & 1K    & 1M    & 100\% & 0.738 & \underline{0.736} & 1.01  & 0.767 & \underline{0.695} & 0.700 & 1.37  & 4.09  & 1.52  & 1.70 \\
  \multicolumn{1}{c|}{} & $10^5$ & 100K  & 10K   & 0\%   & \underline{0.731} & 0.733 & 1.38  & 1.10  & \underline{0.686} & 0.688 & 1.20  & 2.28  & 1.45  & 1.04 \\
  \multicolumn{1}{c|}{} & $10^7$ & 10M   & 100   & 0\%   & 1.01  & \underline{0.891} & 1.49  & 1.05  & 0.970 & \underline{0.847} & 1.10  & 2.48  & 1.69  & 1.06 \\
  \multicolumn{1}{c|}{} & $10^9$ & 1B    & 1     & 0\%   & \underline{0.999} & 1.05  & 1.65  & 1.10  & \underline{0.954} & 1.33  & 1.10  & 2.67  & 1.41  & 1.15 \\
  \multicolumn{1}{c|}{} & \textbf{Avg.} & -     & -     & -     & 0.819 & \underline{0.806} & 1.32  & 0.969 & \underline{0.772} & 0.800 & 1.29  & 2.80  & 1.46  & 1.52 \\
  \midrule
  \multicolumn{1}{c|}{\multirow{6}[2]{*}{\begin{sideways}\textbf{Exponential}\end{sideways}}} & $1\times 10^{-4}$ & 182K  & 100K  & 89.6\% & 0.724 & \underline{0.719} & 1.30  & 0.974 & 0.686 & \underline{0.678} & 1.88  & 2.37  & 1.51  & 1.07 \\
  \multicolumn{1}{c|}{} & $7\times 10^{-5}$ & 252K  & 70.0K & 85.2\% & 0.726 & \underline{0.718} & 1.32  & 0.971 & 0.692 & \underline{0.682} & 1.69  & 2.38  & 1.46  & 1.07 \\
  \multicolumn{1}{c|}{} & $5\times 10^{-5}$ & 343K  & 50.0K & 79.3\% & 0.732 & \underline{0.720} & 1.37  & 1.03  & 0.692 & \underline{0.687} & 1.53  & 2.38  & 1.39  & 1.04 \\
  \multicolumn{1}{c|}{} & $2\times 10^{-5}$ & 789K  & 20.0K & 48.2\% & 0.763 & \underline{0.746} & 1.35  & 1.20  & 0.715 & \underline{0.701} & 1.21  & 2.44  & 1.40  & 1.11 \\
  \multicolumn{1}{c|}{} & $1\times 10^{-5}$ & 1.47M & 10.0K & 0.00\% & 0.815 & \underline{0.782} & 1.37  & 1.18  & 0.756 & \underline{0.708} & 1.13  & 2.49  & 1.38  & 1.11 \\
  \multicolumn{1}{c|}{} & \textbf{Avg.} & -     & -     & -     & 0.751 & \underline{0.737} & 1.34  & 1.06  & 0.708 & \underline{0.691} & 1.46  & 2.41  & 1.43  & 1.08 \\
  \midrule
  \multicolumn{1}{c|}{\multirow{6}[2]{*}{\begin{sideways}\textbf{Zipfian}\end{sideways}}} & 1.5   & 1.79M & 383M  & 97.7\% & 0.657 & \underline{0.642} & 2.30  & 1.62  & 0.614 & \underline{0.607} & 2.31  & 2.49  & 1.71  & 6.59 \\
  \multicolumn{1}{c|}{} & 1.2   & 34.7M & 181M  & 83.6\% & 0.770 & \underline{0.746} & 1.62  & 1.24  & 0.672 & \underline{0.656} & 1.87  & 2.41  & 1.68  & 3.31 \\
  \multicolumn{1}{c|}{} & 1     & 210M  & 46.9M & 42.2\% & 0.909 & \underline{0.892} & 1.32  & 1.09  & 0.802 & \underline{0.761} & 1.38  & 2.60  & 1.46  & 1.74 \\
  \multicolumn{1}{c|}{} & 0.8   & 525M  & 3.22M & 5.32\% & \underline{1.00} & 1.01  & 1.58  & 1.09  & \underline{0.922} & 0.923 & 1.13  & 2.71  & 1.37  & 1.11 \\
  \multicolumn{1}{c|}{} & 0.6   & 756M  & 100K  & 0.10\% & \underline{0.994} & 1.04  & 1.65  & 1.11  & \underline{0.947} & 0.983 & 1.10  & 2.70  & 1.35  & 1.11 \\
  \multicolumn{1}{c|}{} & \textbf{Avg.} & -     & -     & -     & 0.855 & \underline{0.852} & 1.67  & 1.21  & 0.780 & \underline{0.772} & 1.49  & 2.58  & 1.51  & 2.16 \\
  \midrule
  \multicolumn{5}{r|}{\textbf{Overall Geometric Mean}} & 0.807 & \underline{0.797} & 1.44  & 1.08  & \underline{0.753} & 0.753 & 1.41  & 2.59  & 1.46  & 1.53 \\
  \end{tabular}%


  \vspace{-1.5em}
  \caption{\small \textbf{Running times with different input distribution with {$\boldsymbol{n=10^9}$, 64-bit keys and 64-bit values.}}\label{tab:distribution}
    Underlined numbers are the fastest running time in each distribution-input type instance.
    ``parameter'' $=$ distribution parameters (i.e., $\mu$ in uniform, $\lambda$ in exponential, and $s$ in zipfian distribution).
    ``Distinct keys'', ``maximum frequency'', and ``heavy frequency'' are statistics for each test (see details in \cref{sec:exp}).
    \hide{
    ``Ours$_=$'' $=$ our semisort-equal algorithm.
    ``Ours$_<$'' $=$ our semisort-less algorithm.
    ``Ours-i$_{=}$'' and ``Ours-i$_{<}$'' are the corresponding versions for integer key types.
    ``PLSS'' $=$ ParlayLib sample sort~\cite{blelloch2020parlaylib}.
    ``IPS$^4$o'' $=$ IPS$^4$o sample sort~\cite{axtmann2022engineering}.
    ``PLIS'' $=$ ParlayLib integer sort~\cite{blelloch2020parlaylib}.
    ``GSSB'' $=$ GSSB semisort~\cite{gu2015top}.
    ``RS'' $=$ RegionsSort~\cite{obeya2019theoretically}
    ``IPS$^2$Ra'' $=$ IPS$^2$Ra integer sort~\cite{axtmann2022engineering}.
    }
    The algorithm names are described in \cref{tab:baseline}.
    ``Avg.'' means geometric mean numbers across multiple tests.
    \vspace{-1em}
    }
\end{table*}%

%% file: figures/scale.tex

\newcommand{\subfigvspace}{\vspace{-2.5em}}
\begin{figure*}
     \centering
     \begin{subfigure}[b]{0.35\textwidth}
         \centering
         \includegraphics[width=\textwidth]{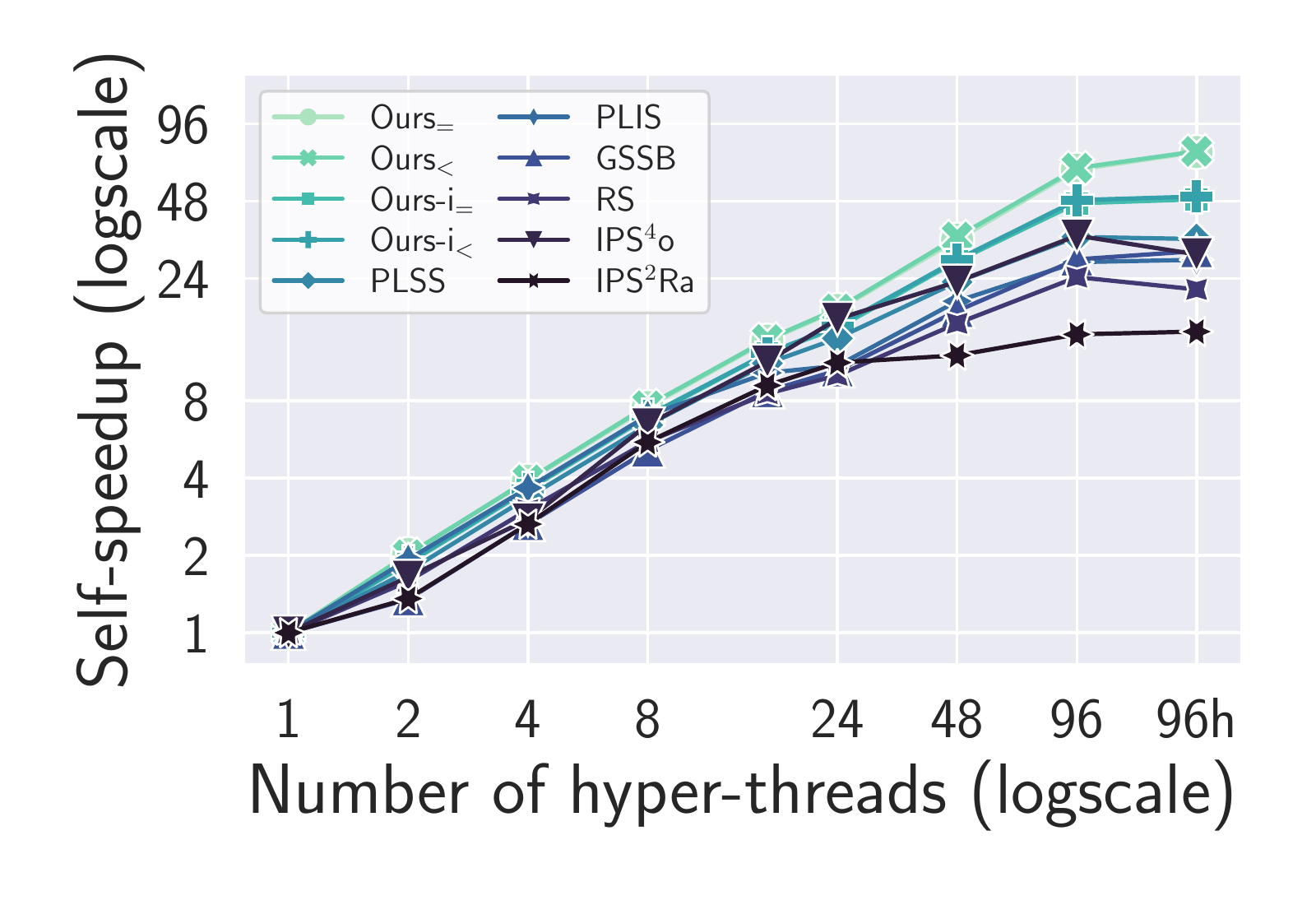}
         \subfigvspace\caption{}
         \label{fig:speedup-zipfian}
     \end{subfigure}
     \hfill
     \begin{subfigure}[b]{0.32\textwidth}
         \centering
         \includegraphics[width=\textwidth]{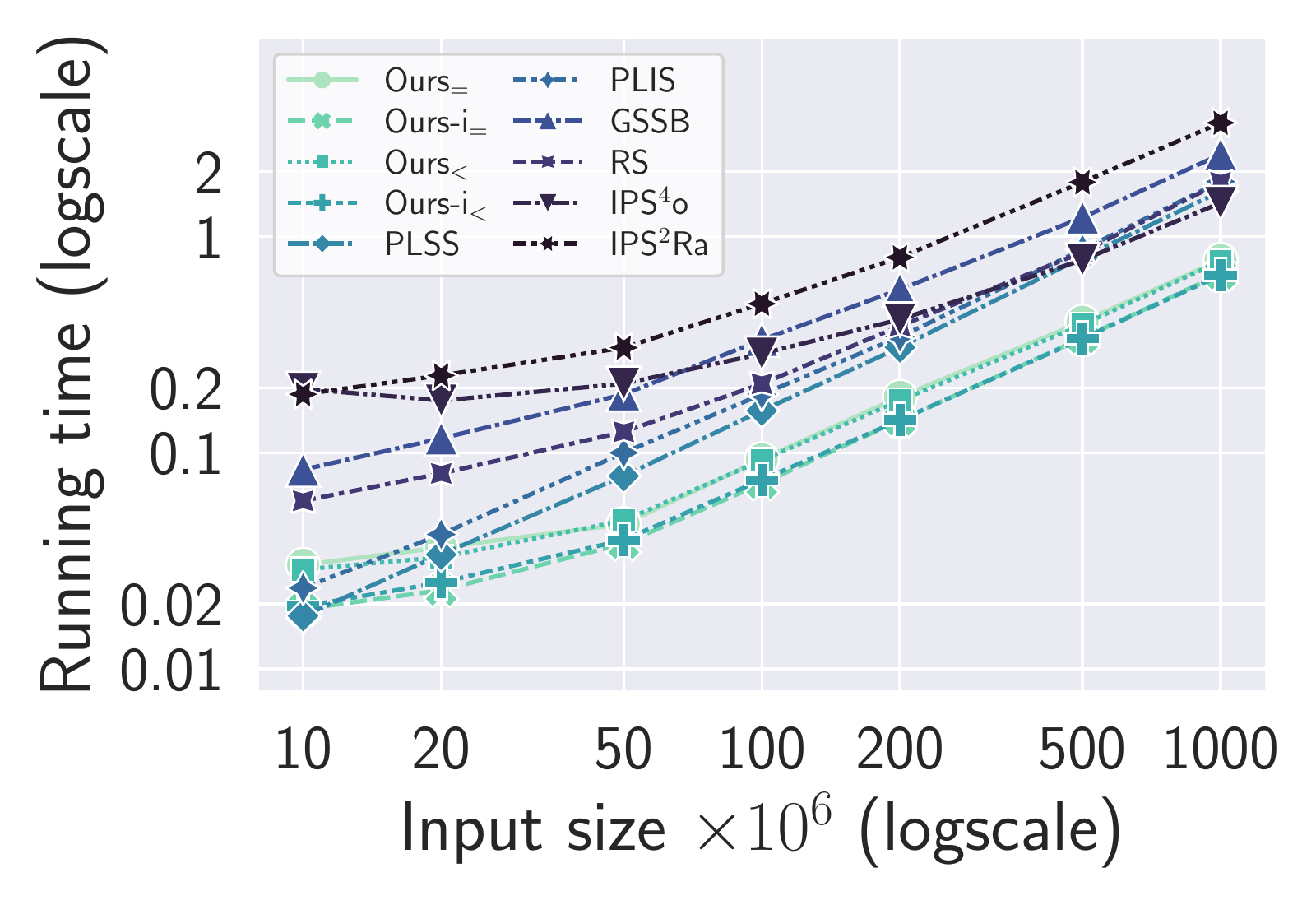}
         \subfigvspace\caption{}
         \label{fig:input-size-scaling-zipfian}
     \end{subfigure}
     \hfill
     \begin{subfigure}[b]{0.31\textwidth}
         \centering
         \includegraphics[width=\textwidth]{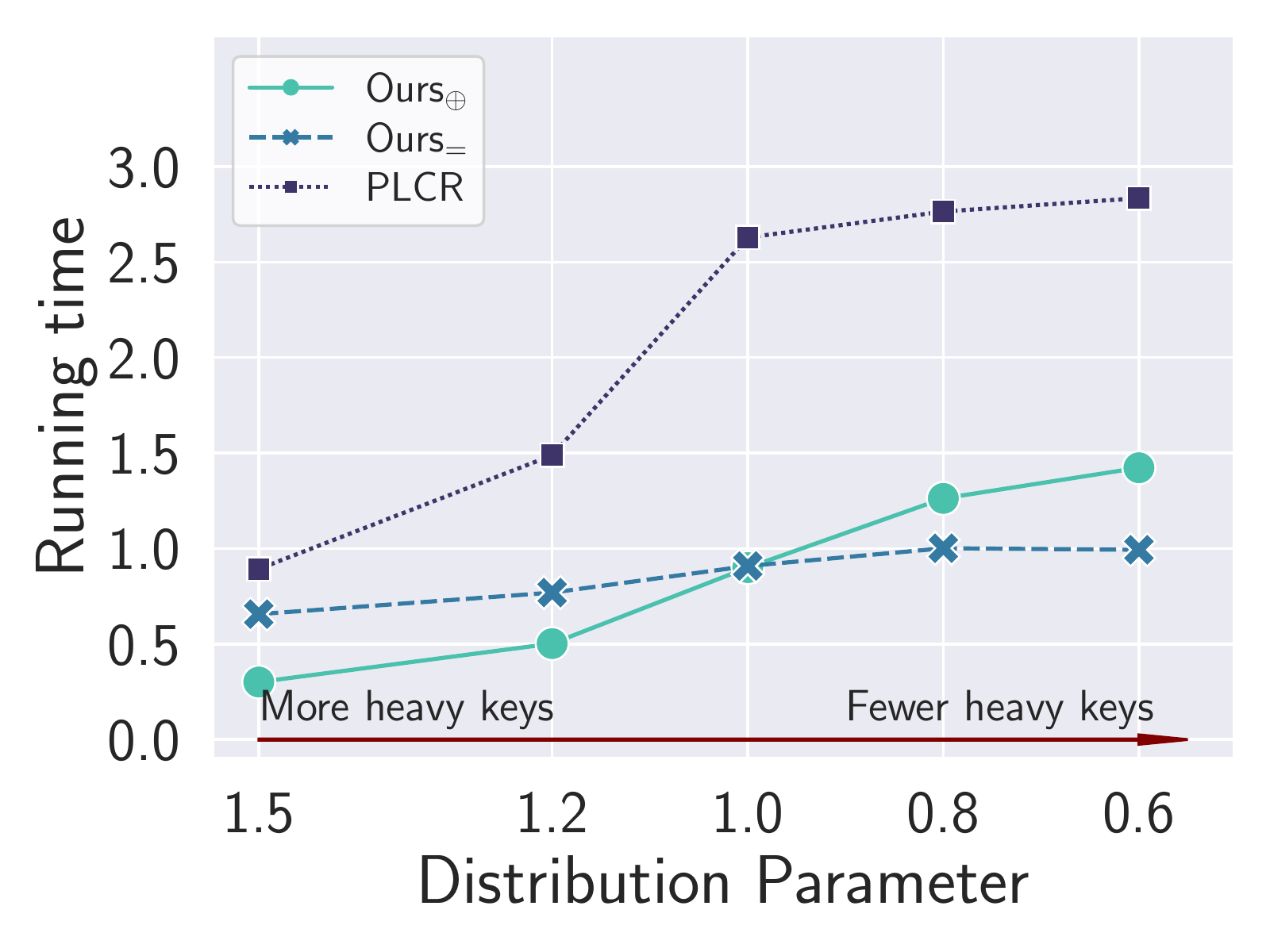}
         \subfigvspace\caption{}
         \label{fig:collect-reduce}
     \end{subfigure}
     \vspace{-2em}
        \caption{\small  \textbf{(a). Self-speedup of all tested implementations with increasing hyper-thread counts on Zipfian-$\boldsymbol{1.2}$.} $n=10^9$.
        \textbf{(b). Scalability with increasing input size ($n$) of all tested implementations on Zipfian-$\boldsymbol{1.2}$.} 
        \textbf{(c). Performance of \collectreduce{} with various Zipfian distributions.}  $n=10^9$. 
    Ours$_{\oplus}$ is our collect-reduce algorithm.
    Ours$_{=}$ is our \semisortequal{} algorithm.
    \plcr is the collect-reduce in ParlayLib~\cite{blelloch2020parlaylib}.
    All three cases are on 64-bit keys and 64-bit values.\label{fig:three graphs}}\vspace{-2em}        
\end{figure*}

%% file: figures/input-type.tex
\begin{figure}[t]
    \centering
    \includegraphics[width=.9\columnwidth]{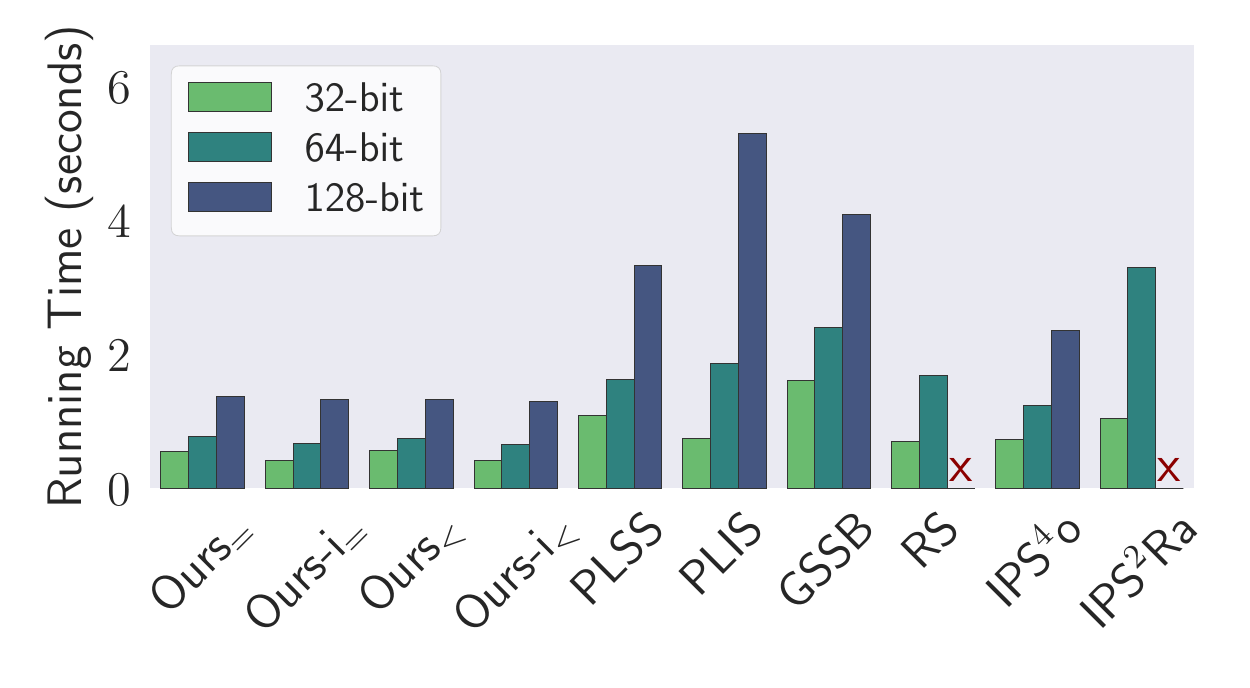}
    \vspace{-1em}
    \caption{\small \textbf{Running time of our semisort implementations and other implementations with different key-lengths on Zipfian-$\boldsymbol{1.2}$.} $n=10^9$. We put crosses on RS and IPS$^2$Ra because they do not support 128-bit keys.\label{fig:scaling-keylengths-zipfian}}\vspace{-1em}
\end{figure}

%% file: tables/applications.tex
\begin{table*}
\begin{tabular}{rrrrrr|rrrrrrrr}
  \multicolumn{1}{r|}{} & \multicolumn{1}{l}{\boldmath{}\textbf{$n$}\unboldmath{}} & \multicolumn{1}{l}{\boldmath{}\textbf{$m$}\unboldmath{}} & \multicolumn{1}{c}{\boldmath{}\textbf{$n_{\mathit{dist}}$}\unboldmath{}} & \multicolumn{1}{c}{\boldmath{}\textbf{$f_\mathit{max}$}\unboldmath{}} & \multicolumn{1}{c|}{\boldmath{}\textbf{$r_\mathit{heavy}$}\unboldmath{}} & \multicolumn{1}{l}{\boldmath{}\textbf{Ours-i$_=$}\unboldmath{}} & \multicolumn{1}{l}{\boldmath{}\textbf{Ours-i$_<$}\unboldmath{}} & \multicolumn{1}{l}{\textbf{PLSS}} & \multicolumn{1}{l}{\boldmath{}\textbf{IPS$^4$o}\unboldmath{}} & \multicolumn{1}{l}{\textbf{PLIS}} & \multicolumn{1}{l}{\textbf{GSSB}} & \multicolumn{1}{l}{\textbf{RS}} & \multicolumn{1}{l}{\boldmath{}\textbf{IPS$^2$Ra}\unboldmath{}} \\
  \hline
  \multicolumn{1}{l|}{\textbf{LJ~\cite{backstrom2006group}}} & 4.85M & 69.0M & 4.49M & 13.9K & 62.8K & 0.042 & 0.045 & 0.075 & 0.101 & \underline{0.039} & 4.56  & 0.062 & s.g. \\
  \multicolumn{1}{l|}{\textbf{TW~\cite{kwak2010twitter}}} & 41.7M & 1.47B & 35.7M & 770K  & 74.8M & \underline{0.714} & 0.834 & 1.57  & 0.814 & 0.900 & t.o.     & 1.06  & 2.94 \\
  \multicolumn{1}{l|}{\textbf{CM~\cite{cosmo50,wang2021geograph}}} & 321M  & 1.61B & 320M  & 17    & 0     & \underline{0.791} & 1.04  & 1.84  & 1.10  & 0.903 & 3.58  & 1.09  & 1.44 \\
  \multicolumn{1}{l|}{\textbf{SD~\cite{webgraph}}} & 89.2M & 2.04B & 72.8M & 2.34M & 456M  & \underline{0.916} & 1.08  & 2.10  & 1.16  & 1.24  & s.g.     & 1.37  & 2.82 \\
  \hline
  \multicolumn{6}{r|}{\textbf{Overall geometric mean}} & \underline{0.385} & 0.452 & 0.821 & 0.569 & 0.446 & -     & 0.559 & - \\
  \end{tabular}%

  \vspace{-.15in}
  \caption{\small\textbf{Running time on graph transposing (in seconds). }
  $n=$ number of vertices.
  $m=$ number of edges.
  $n_\mathit{dist}=$ number of distinct keys. $f_\mathit{max}=$ maximum frequency.
  $r_\mathit{heavy}=$ ratio of keys with more than $500\log n$ occurrences.
  ``t.o.'' $=$ did not finish in one minute.
  ``s.g.'' $=$ segmentation fault.
  \label{tab:graph_transposing}%
  }
  \vspace{-.5em}
\end{table*}


\begin{table}
\small
\begin{tabular}{@{}r@{ }r@{ }r@{ }r@{ }r@{ }|r@{ }rrr@{}}
  & \multicolumn{1}{l}{\boldmath{}\textbf{$n$}\unboldmath{}} & \multicolumn{1}{c}{\boldmath{}\textbf{$n_{\mathit{dist}}$}\unboldmath{}} & \multicolumn{1}{c}{\boldmath{}\textbf{$f_\mathit{max}$}\unboldmath{}} & \multicolumn{1}{@{ }c@{ }|}{\boldmath{}\textbf{$r_\mathit{heavy}$}\unboldmath{}} & \multicolumn{1}{l@{ }}{\boldmath{}\textbf{Ours$_=$}\unboldmath{}} & \multicolumn{1}{@{ }l@{ }}{\boldmath{}\textbf{Ours$_<$}\unboldmath{}} & \multicolumn{1}{l@{  }}{\textbf{PLSS}} & \multicolumn{1}{@{ }l@{}}{\boldmath{}\textbf{IPS$^4$o}\unboldmath{}} \\
\hline
\multicolumn{1}{@{}l}{\textbf{2-gram}} & 68.0M & 3.12M & 2.18M & 28.0\% & \underline{0.312} & 0.332 & 0.346 & 0.753 \\
\multicolumn{1}{@{}l}{\textbf{3-gram}} & 224M  & 47.5M & 319K  & 4.43\% & \underline{1.44} & 1.80  & 2.00  & 3.26 \\
\hline
\multicolumn{5}{r|}{\textbf{Overall geometric mean}} & \underline{0.671} & 0.772 & 0.832 & 1.57 \\
\end{tabular}%

  \vspace{-.2in}
  \caption{\small \textbf{Running time on semisorting n-grams~\cite{wikipediangram} (in seconds). }
  $n=$ number of records.
  $n_\mathit{dist}=$ number of distinct keys.
  $f_\mathit{max}=$ maximum frequency.
  $r_\mathit{heavy}=$ ratio of keys with more than $500\log n$ occurrences.
  \label{fig:ngrams}}\vspace{-.5em}
\end{table}

%% file: conclusion.tex
\section{Conclusions and Future Work}\label{sec:conclusion}
In this paper, we designed flexible and high-performance algorithms for semisort and related problems.
We presented two implementations, \semisortequal{} (only the equality-test is required),
and \semisortless{} (the less-than-test is also available).
Compared to previous semisort algorithms, our new algorithms yield
improvements in terms of  space-efficiency and I/O-friendliness,
ensure stability and determinism, and importantly, increase the
flexibility of the interface.
On different input distributions, input sizes and key lengths, our implementations
achieve high performance, and outperform existing sorting and semisorting algorithms in most of the tests.
For example, on $10^9$ 64-bit keys,
on all the tested distributions, (one of) our algorithms are always the fastest among all tested algorithms,
and the other one always performs similarly.

Based on our experiments, in-place versions of the sorting algorithms (e.g., \ipso{}) are competitive and sometimes more efficient than the non-in-place versions (e.g., \plss{}).
The good performance for the in-place algorithms is due to the I/O savings in the distributing step---they use the same array for both the input and the buckets ($A$ and $T$ in \cref{alg:semisort}).
We note that the new techniques proposed in this paper are independent of this distribution step.
An interesting future direction is to redesign this step (e.g., borrowing ideas from \ipso{}) to improve the overall performance and reduce the extra space usage.

\section*{Acknowledgement}

This work is supported by NSF grants CCF-2103483, IIS-2227669, NSF CAREER award CCF-2238358, and UCR Regents Faculty Fellowships.
We thank anonymous reviewers for the useful feedbacks.

%% file: appendix.tex
\begin{figure}[!h]
    \centering
    \includegraphics[width=\columnwidth]{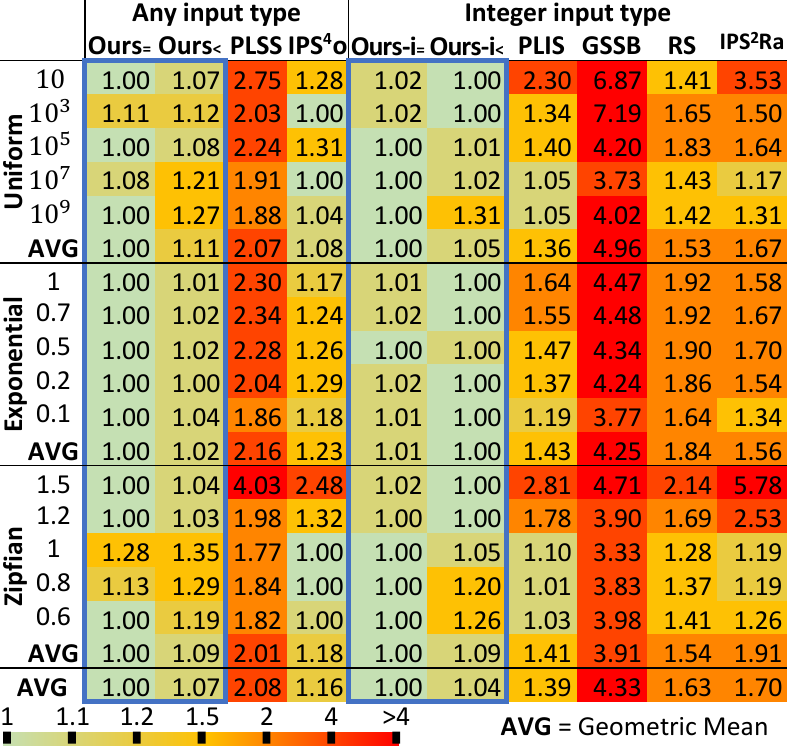}
    \caption{\textbf{Heatmap of the relative performance of all implementations normalized to the fastest in each test.} $n=10^9$. \textbf{32-bit keys and 32-bit values}.
    The parameters in exponential distributions are multiplied by $10^{4}$.}\label{fig:heatmap-32}
  \end{figure}

\begin{figure}[!h]
  \centering
  \includegraphics[width=\columnwidth]{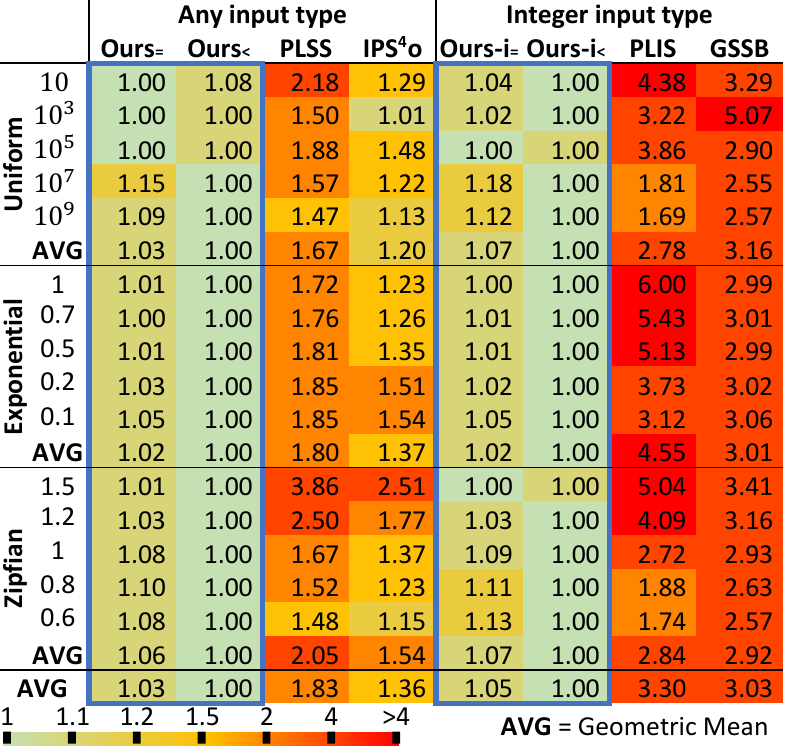}
  \caption{\textbf{Heatmap of the relative performance of all implementations normalized to the fastest in each test.} $n=10^9$. \textbf{128-bit keys and 128-bit values.}
  The parameters in exponential distributions are multiplied by $10^{4}$.
  RS and IPS$^2$Ra do not support 128-bit keys, so we remove them from this heatmap.}\label{fig:heatmap-128}
\end{figure}

\begin{figure*}[!h]
{\Large Self-speedup with Varying Thread Counts ($n=10^9$, 64-bit keys)}\\
Uniform Distribution\\
\begin{minipage}{.95\columnwidth}
  \centering
  \includegraphics[width=\columnwidth]{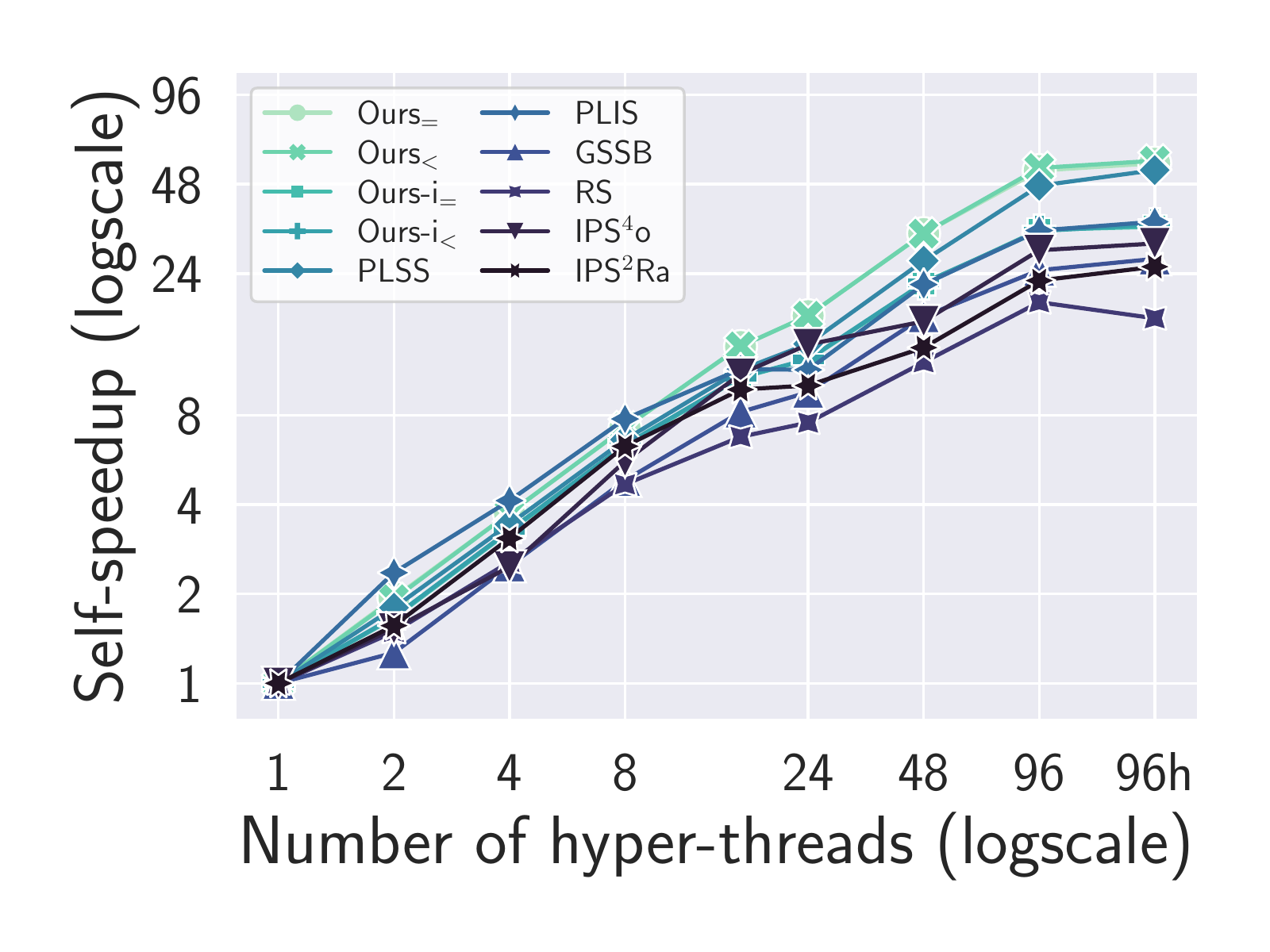}
  \caption{\textbf{Self-speedup with varying thread counts of all tested implementations on uniform-$\boldsymbol{10^3}$.}}\label{fig:scalability-uniform-1e3}
\end{minipage}\hfil
\begin{minipage}{.95\columnwidth}
  \centering
  \includegraphics[width=\columnwidth]{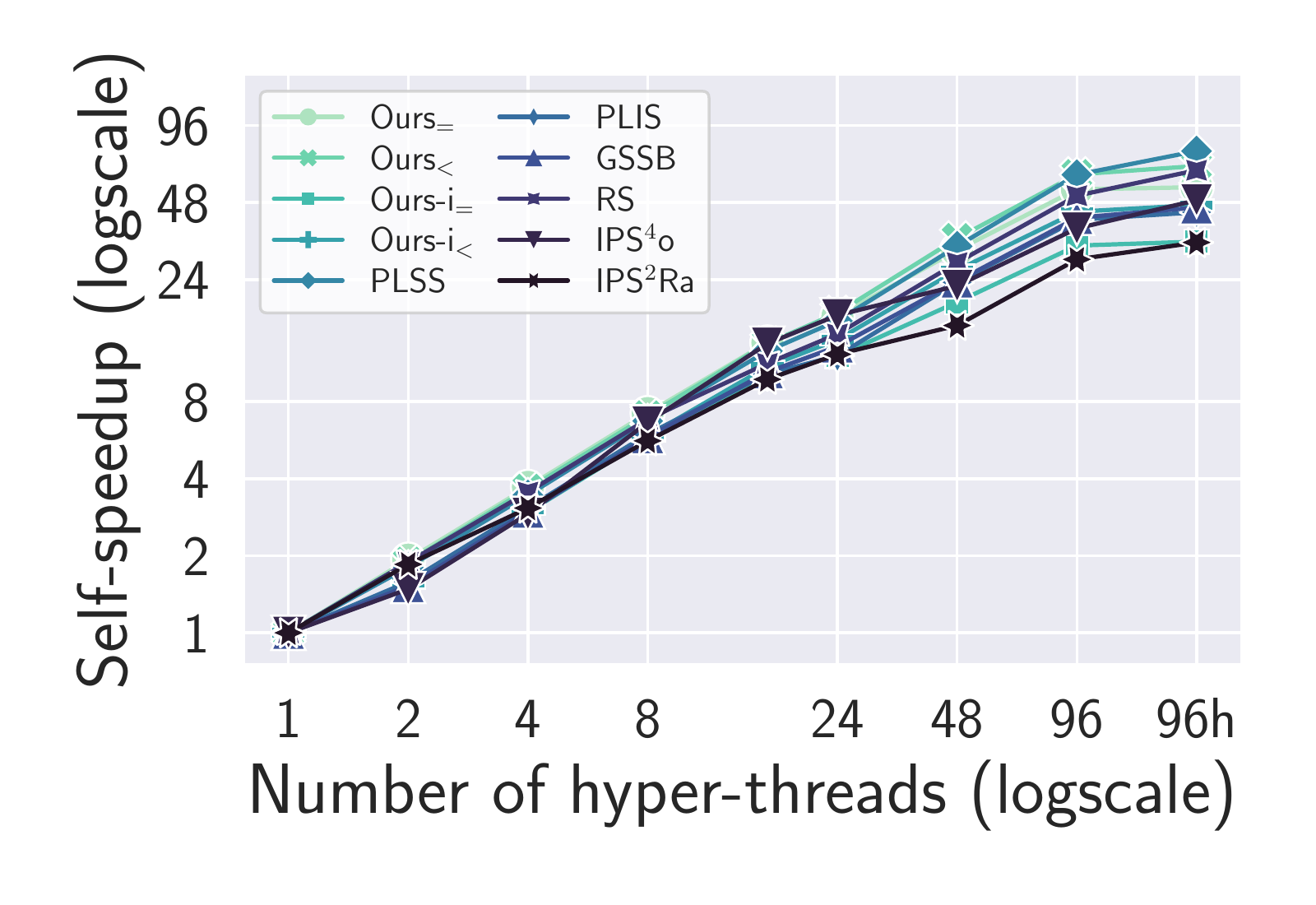}
  \caption{\textbf{Self-speedup with varying thread counts of all tested implementations on uniform-$\boldsymbol{10^7}$.}}\label{fig:scalability-uniform-1e7}
\end{minipage}

Exponential Distribution\\
\begin{minipage}{.95\columnwidth}
  \centering
  \includegraphics[width=\columnwidth]{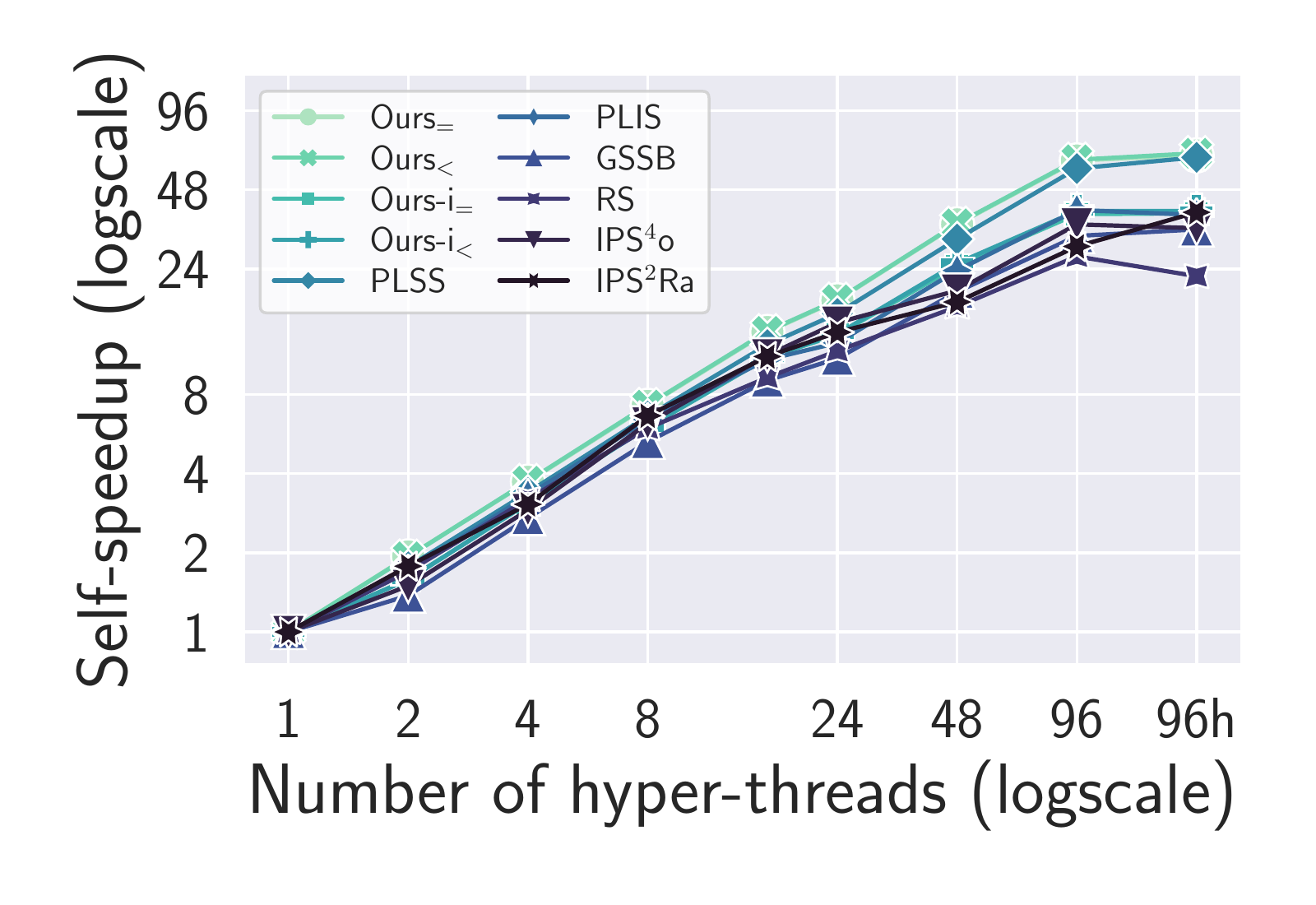}
  \caption{\textbf{Self-speedup with varying thread counts of all tested implementations on exponential-$\boldsymbol{.00002}$.}}\label{fig:scalability-exponential-2e-5}
\end{minipage}\hfil
\begin{minipage}{.95\columnwidth}
  \centering
  \includegraphics[width=\columnwidth]{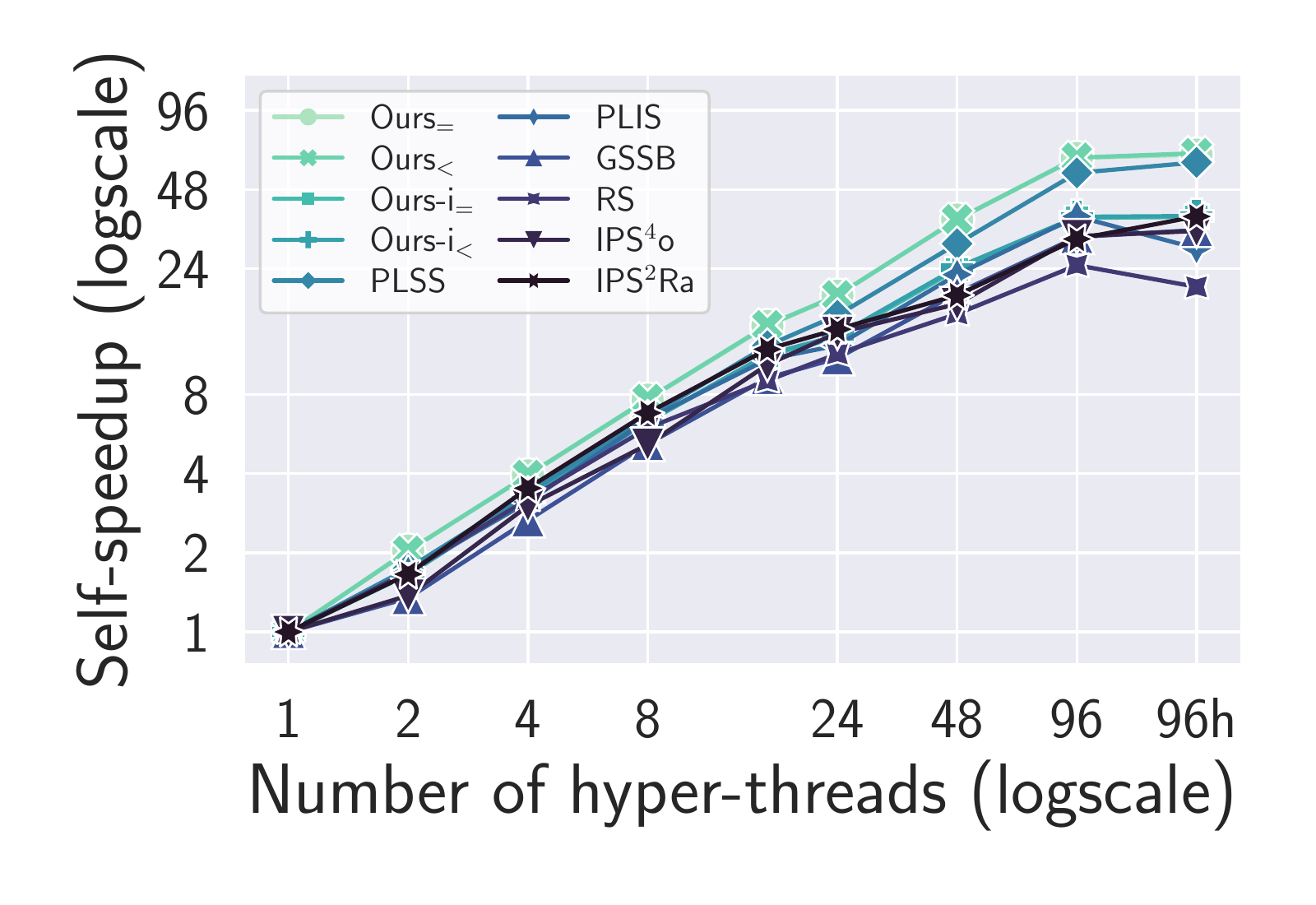}
  \caption{\textbf{Self-speedup with varying thread counts of all tested implementations on exponential-$\boldsymbol{.00007}$.}}\label{fig:scalability-exponential-7e-5}
\end{minipage}

Zipfian Distribution\\
\begin{minipage}{.95\columnwidth}
  \centering
  \includegraphics[width=\columnwidth]{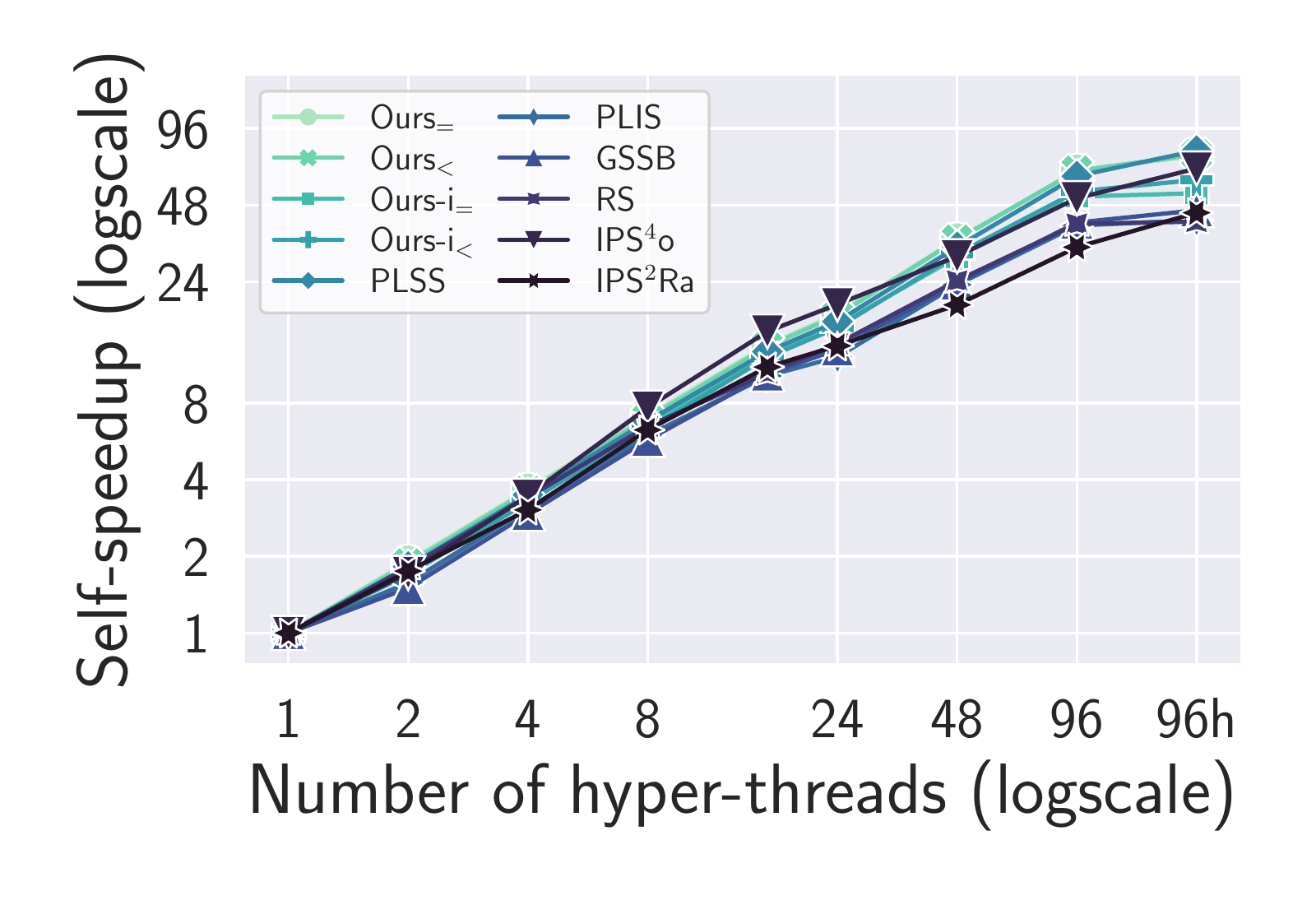}
  \caption{\textbf{Self-speedup with varying thread counts of all tested implementations on Zipfian-$\boldsymbol{0.8}$.}}\label{fig:scalability-Zipfian-0.8}
\end{minipage}\hfil
\begin{minipage}{.95\columnwidth}
  \centering
  \includegraphics[width=\columnwidth]{figures/output/scaling/zipfian-1.2.pdf}
  \caption{\textbf{Self-speedup with varying thread counts of all tested implementations on Zipfian-$\boldsymbol{1.2}$.}}\label{fig:scalability-Zipfian-1.2}
\end{minipage}
\end{figure*}

\begin{figure*}[!h]
{\Large Varying Input Sizes (64-bit keys)}\\~\\

Uniform Distribution\\
\begin{minipage}{.9\columnwidth}
  \centering
  \includegraphics[width=\columnwidth]{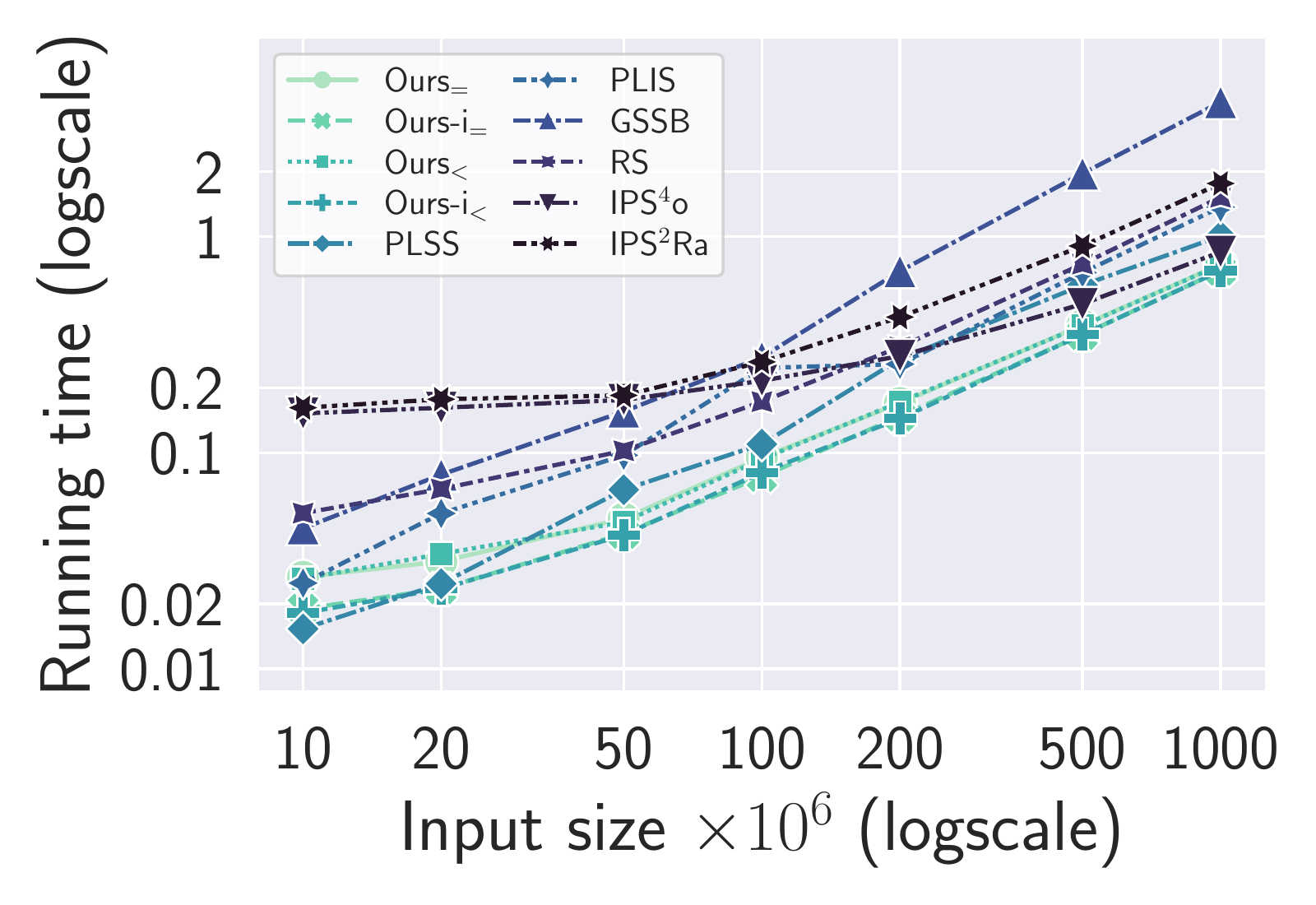}
  \caption{\textbf{Scalability with increasing input size ($n$) of all tested implementations on uniform-$\boldsymbol{10^3}$.}}\label{fig:input-size-uniform-1e3}
\end{minipage}\hfil
\begin{minipage}{.9\columnwidth}
  \centering
  \includegraphics[width=\columnwidth]{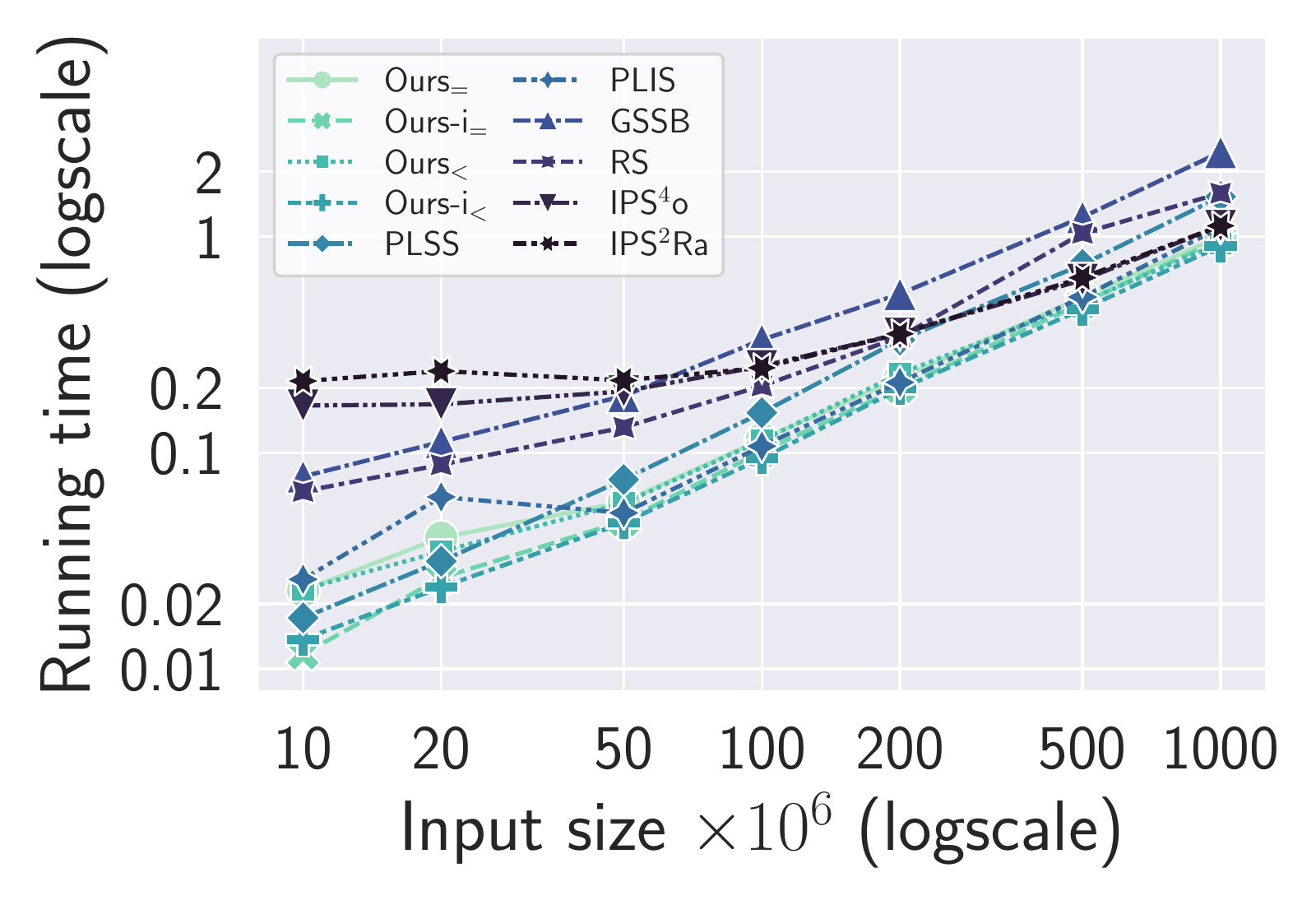}
  \caption{\textbf{Scalability with increasing input size ($n$) of all tested implementations on uniform-$\boldsymbol{10^7}$.}}\label{fig:input-size-uniform-1e7}
\end{minipage}

Exponential Distribution\\
\begin{minipage}{.9\columnwidth}
  \centering
  \includegraphics[width=\columnwidth]{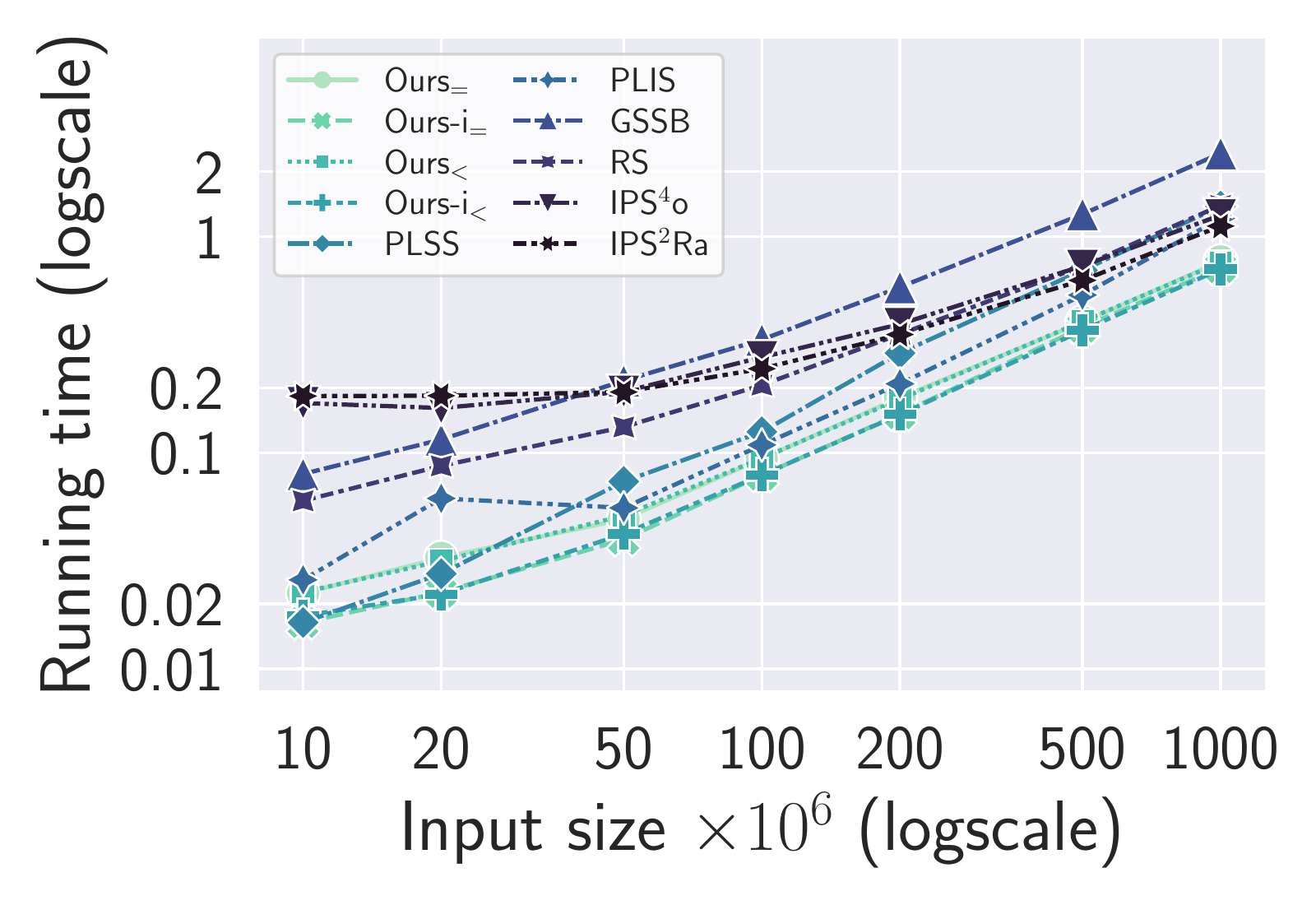}
  \caption{\textbf{Scalability with increasing input size ($n$) of all tested implementations on exponential-$\boldsymbol{.00002}$.}}\label{fig:input-size-exponential-2e-5}
\end{minipage}\hfil
\begin{minipage}{.9\columnwidth}
  \centering
  \includegraphics[width=\columnwidth]{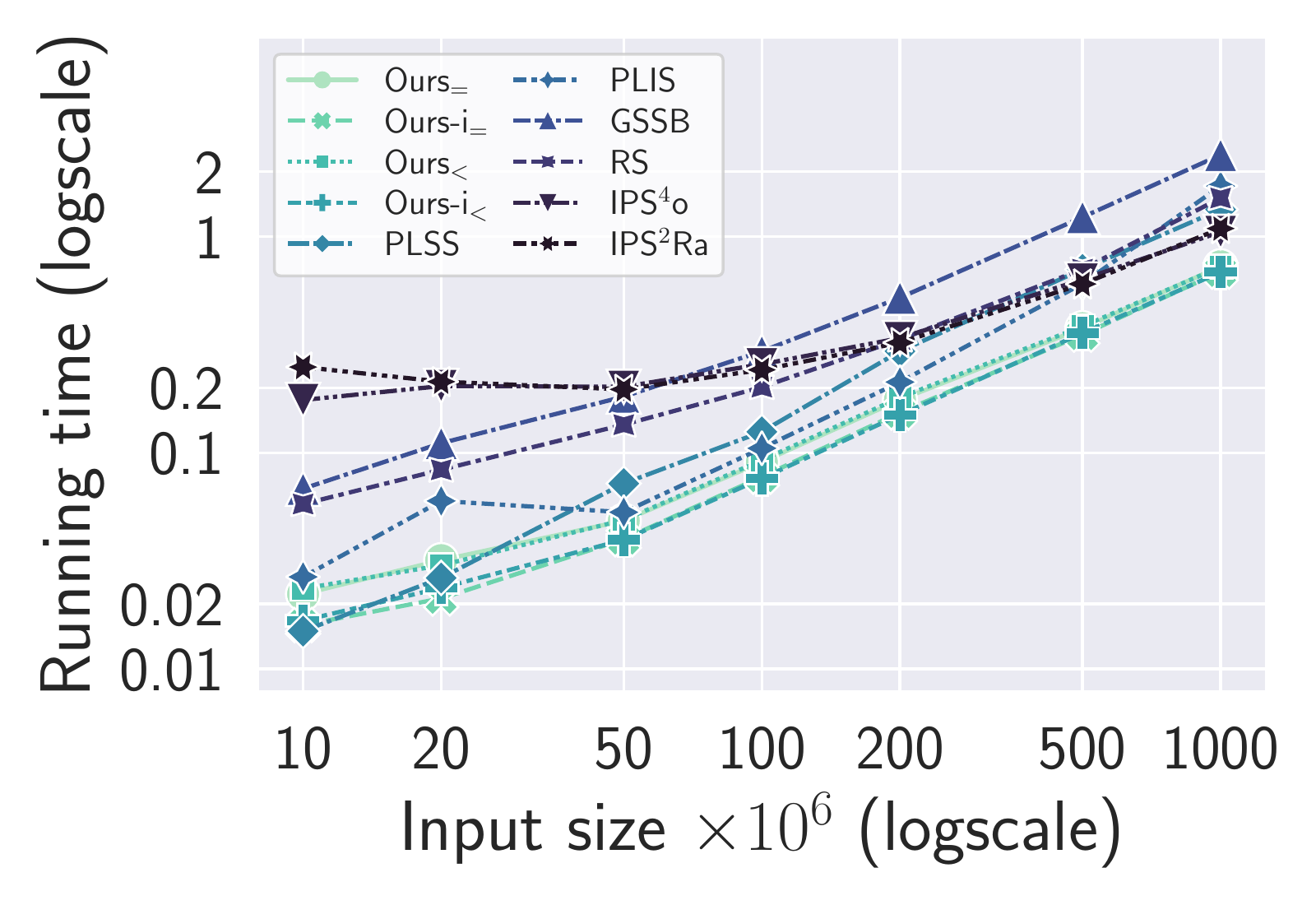}
  \caption{\textbf{Scalability with increasing input size ($n$) of all tested implementations on exponential-$\boldsymbol{.00007}$.}}\label{fig:input-size-exponential-7e-5}
\end{minipage}

Zipfian Distribution\\
\begin{minipage}{.9\columnwidth}
  \centering
  \includegraphics[width=\columnwidth]{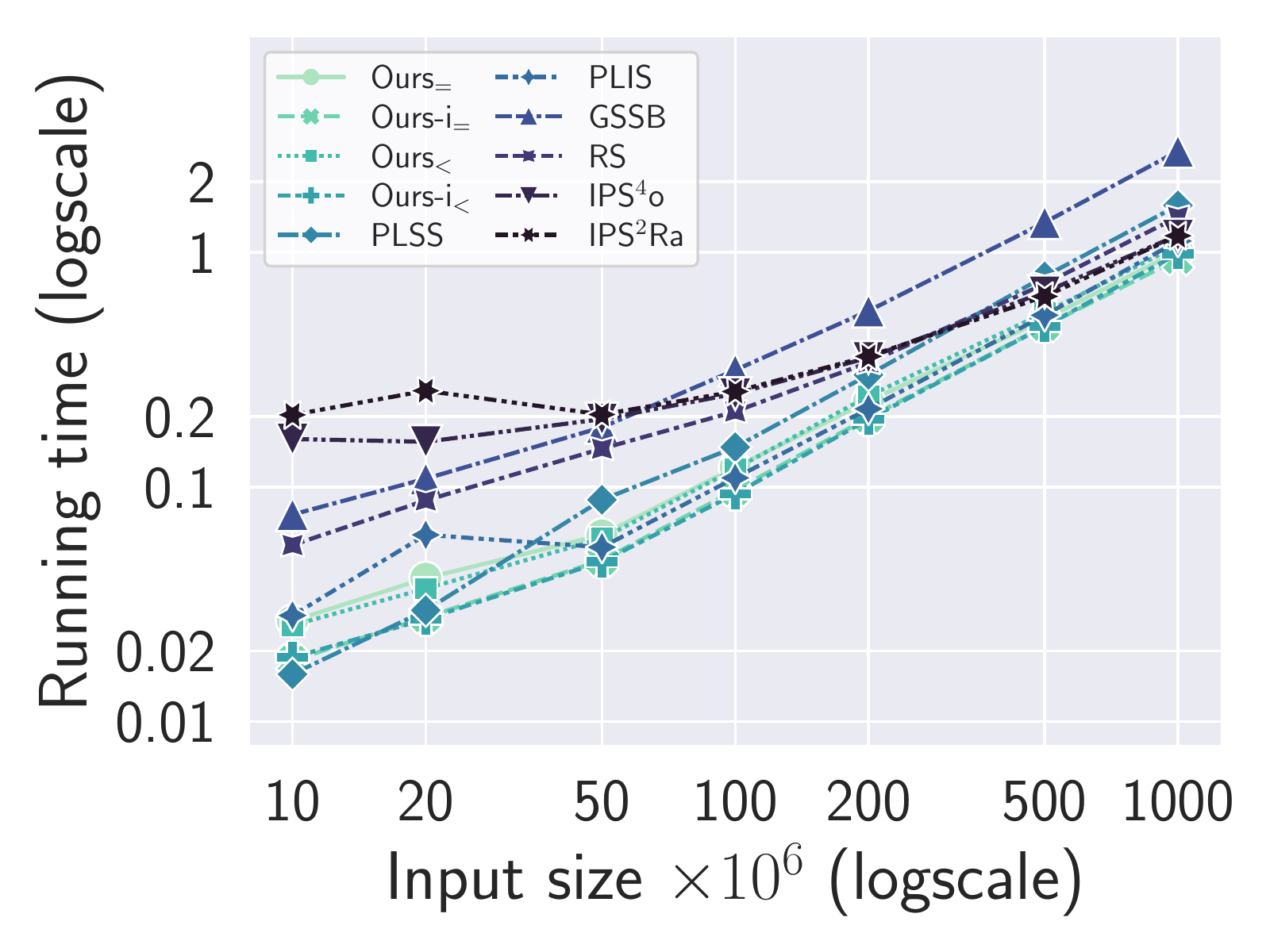}
  \caption{\textbf{Scalability with increasing input size ($n$) of all tested implementations on Zipfian-$\boldsymbol{0.8}$.}}\label{fig:input-size-Zipfian-0.8}
\end{minipage}\hfil
\begin{minipage}{.9\columnwidth}
  \centering
  \includegraphics[width=\columnwidth]{figures/output/increasing_input_size/zipfian-1.2.pdf}
  \caption{\textbf{Scalability with increasing input size ($n$) of all tested implementations on Zipfian-$\boldsymbol{1.2}$.}}\label{fig:input-size-Zipfian-1.2}
\end{minipage}
\end{figure*}

\begin{figure*}[!h]
{\Large Scalability with Varying Key Lengths ($n=10^9$, 64-bit keys)}\\~\\
Uniform Distribution\\
\begin{minipage}{.95\columnwidth}
  \centering
  \includegraphics[width=\columnwidth]{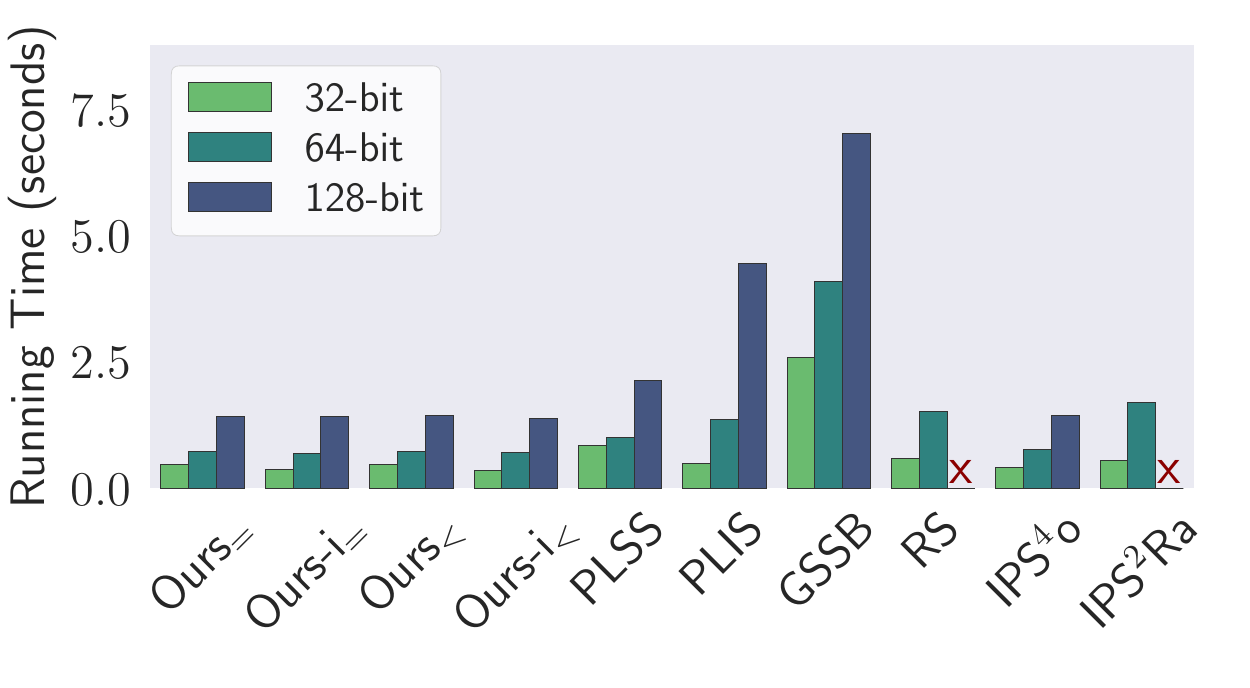}
  \caption{\textbf{Scalability with different key lengths of all tested implementations on uniform-$\boldsymbol{10^3}$.} We put crosses on RS and IPS$^2$Ra because they do not support 128-bit keys.}\label{fig:key-size-uniform-1e3}
\end{minipage}\hfil
\begin{minipage}{.95\columnwidth}
  \centering
  \includegraphics[width=\columnwidth]{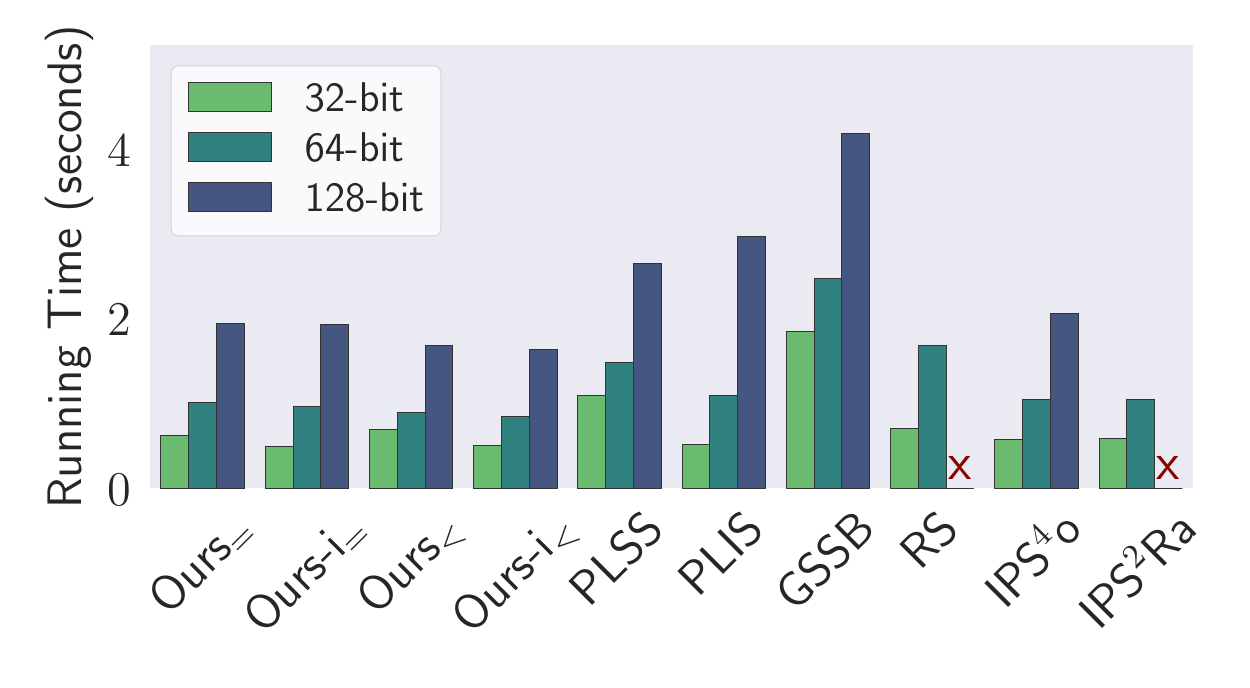}
  \caption{\textbf{Scalability with different key lengths of all tested implementations on uniform-$\boldsymbol{10^7}$.} We put crosses on RS and IPS$^2$Ra because they do not support 128-bit keys.}\label{fig:key-size-uniform-1e7}
\end{minipage}

Exponential Distribution\\
\begin{minipage}{.95\columnwidth}
  \centering
  \includegraphics[width=\columnwidth]{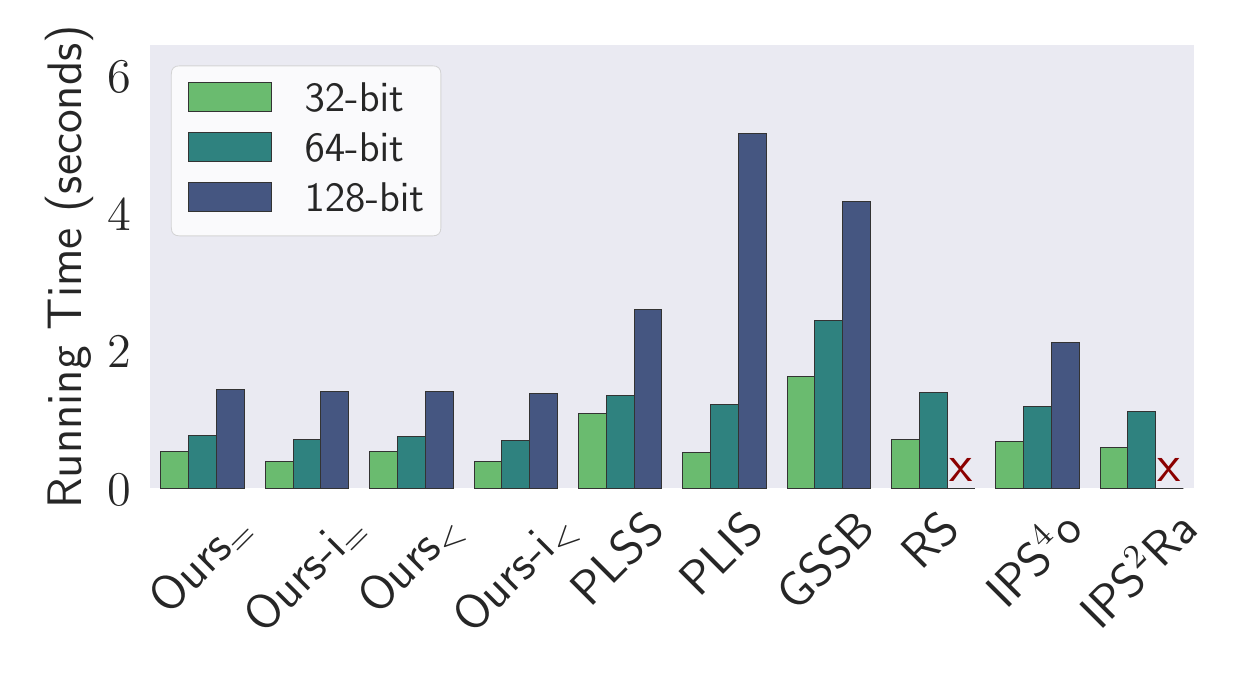}
  \caption{\textbf{Scalability with different key lengths of all tested implementations on exponential-$\boldsymbol{.00002}$.} We put crosses on RS and IPS$^2$Ra because they do not support 128-bit keys.}\label{fig:key-size-exponential-2e-5}
\end{minipage}\hfil
\begin{minipage}{.95\columnwidth}
  \centering
  \includegraphics[width=\columnwidth]{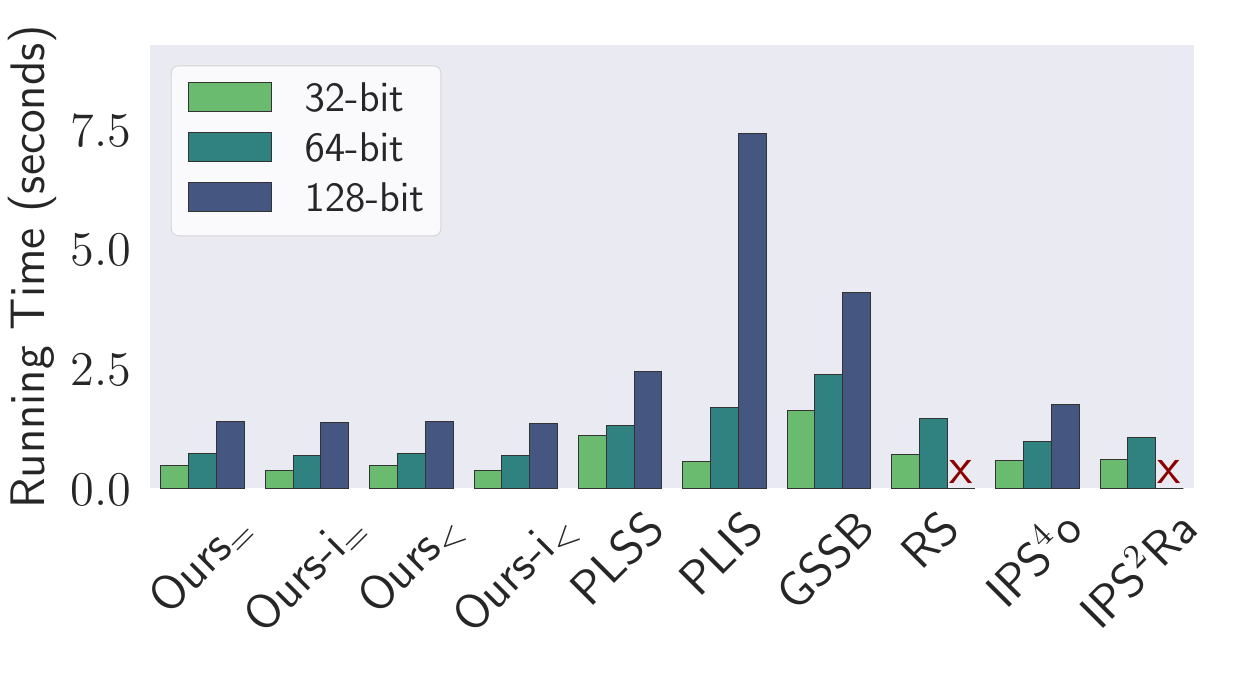}
  \caption{\textbf{Scalability with different key lengths of all tested implementations on exponential-$\boldsymbol{.00007}$.} We put crosses on RS and IPS$^2$Ra because they do not support 128-bit keys.}\label{fig:key-size-exponential-7e-5}
\end{minipage}

Zipfian Distribution\\
\begin{minipage}{.95\columnwidth}
  \centering
  \includegraphics[width=\columnwidth]{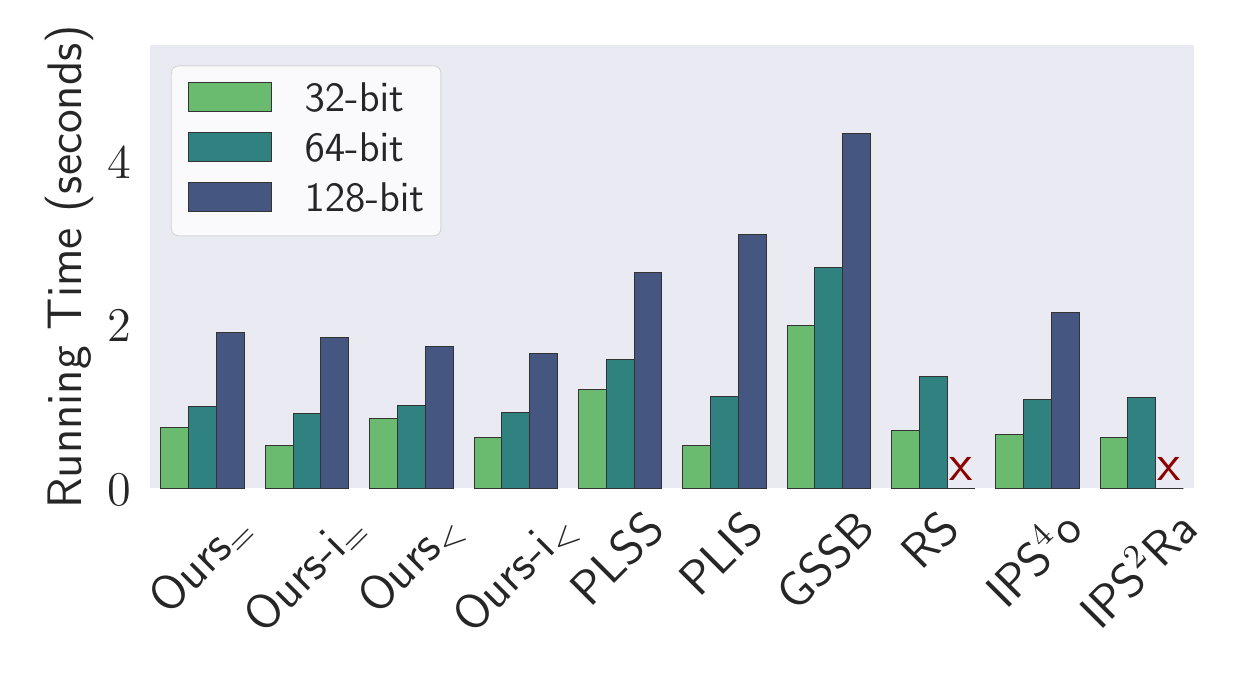}
  \caption{\textbf{Scalability with different key lengths of all tested implementations on Zipfian-$\boldsymbol{0.8}$.} We put crosses on RS and IPS$^2$Ra because they do not support 128-bit keys.}\label{fig:key-size-Zipfian-0.8}
\end{minipage}\hfil
\begin{minipage}{.95\columnwidth}
  \centering
  \includegraphics[width=\columnwidth]{figures/output/vs_key_length/zipfian-1.2.pdf}
  \caption{\textbf{Scalability with different key lengths of all tested implementations on Zipfian-$\boldsymbol{1.2}$.} We put crosses on RS and IPS$^2$Ra because they do not support 128-bit keys.}\label{fig:key-size-Zipfian-1.2}
\end{minipage}
\end{figure*}

\begin{figure*}[!h]
{\Large Performance for Collect-Reduce ($n=10^9$, 64-bit keys)}\\~\\~\\
\begin{minipage}{\columnwidth}
\qquad Uniform Distribution\\
  \centering
  \includegraphics[width=\columnwidth]{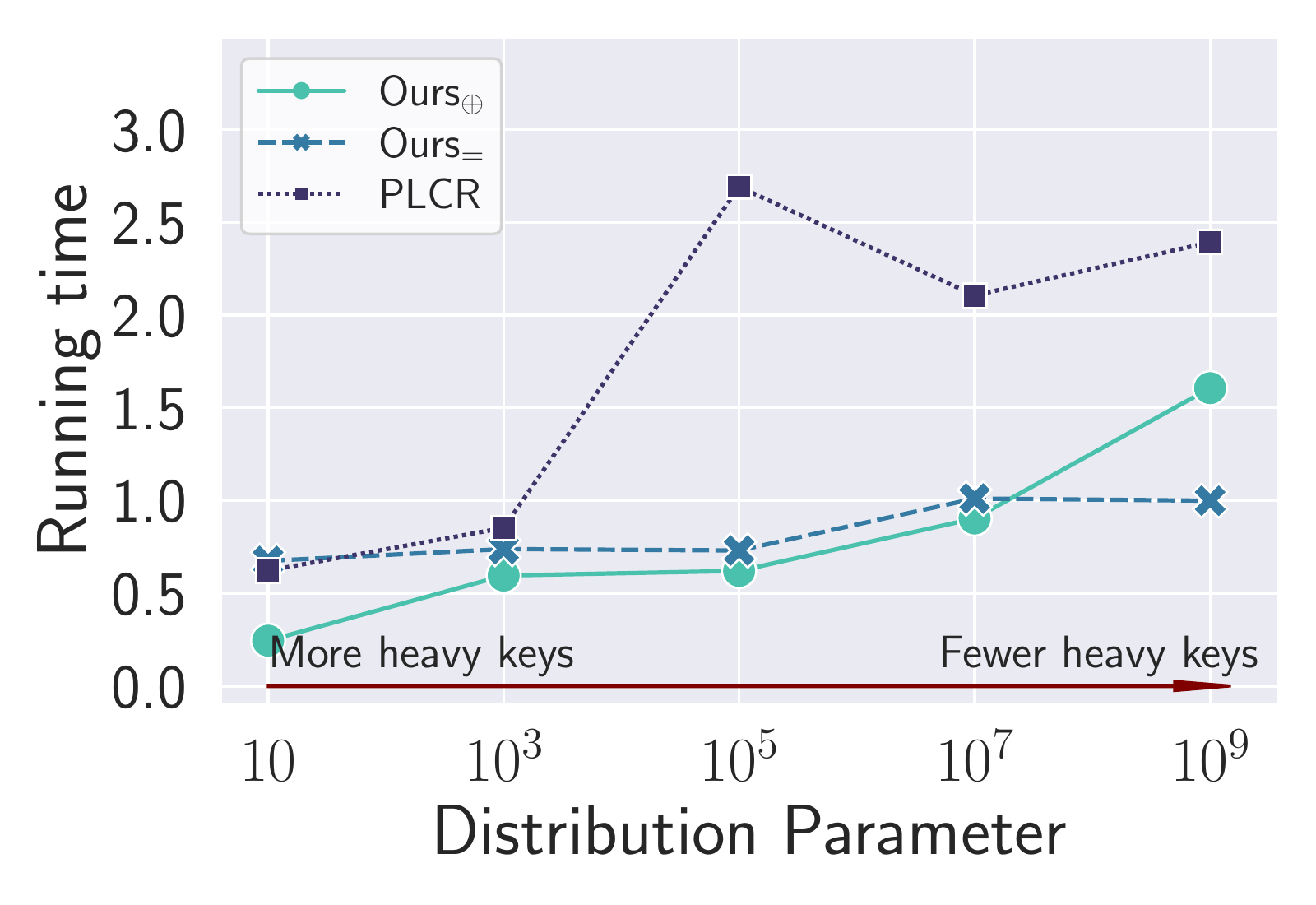}
  \caption{\textbf{Performance comparison between our collect-reduce (Ours$_\oplus$), ParlayLib collect-reduce (PLCR), and \semisortequal (Ours$_=$).}}\label{fig:cr-uni}
\end{minipage}\hfill
\begin{minipage}{\columnwidth}
\qquad Exponential Distribution\\
  \centering
  \includegraphics[width=\columnwidth]{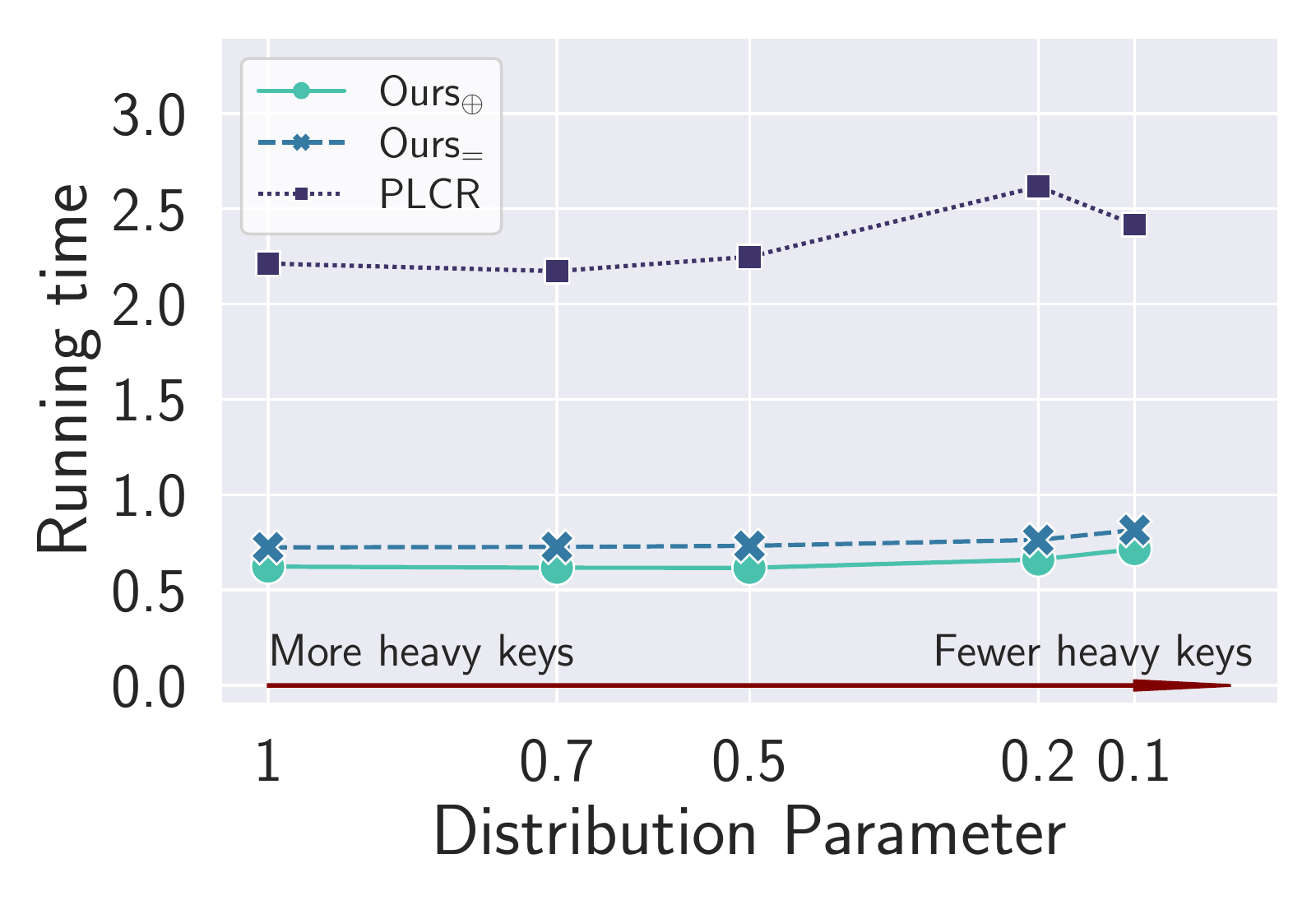}
  \caption{\textbf{Performance comparison between our collect-reduce (Ours$_\oplus$) , ParlayLib collect-reduce (PLCR), and \semisortequal (Ours$_=$).} The parameters are multiplied by $10^{4}$.}\label{fig:cr-exp}
\end{minipage}

\qquad Zipfian Distribution\\
\begin{minipage}{\columnwidth}
  \centering
  \includegraphics[width=\columnwidth]{figures/output/collect_reduce/zipfian.pdf}
  \caption{\textbf{Performance comparison between our collect-reduce (Ours$_\oplus$), ParlayLib collect-reduce (PLCR), and \semisortequal (Ours$_=$).}}\label{fig:cr-zip}
\end{minipage}
\end{figure*}